\documentclass[journal]{IEEEtran}

\usepackage{cite}
\usepackage[dvipsnames]{xcolor}
\usepackage{tikz}
\usetikzlibrary{external}
\usetikzlibrary{positioning}
\usepackage{pgfplots}
\pgfplotsset{compat=1.15}
\usepgfplotslibrary{groupplots}

\usepackage{amsmath}
\usepackage{amssymb}

\usepackage{url}

\usepackage{acro}

\usepackage{array}
\usepackage{booktabs}
\usepackage{calc}

\usepackage[inline]{enumitem}

\tikzset{
    png export/.style={
        external/system call/.add={}{; convert -density 400 "\image.pdf" "\image.png"},
        /pgf/images/external info,
        /pgf/images/include external/.code={
            \includegraphics[width=\pgfexternalwidth,height=\pgfexternalheight]{##1.png}
        },
    }
}

\pgfplotsset{%
    evalline/.style={line width=1pt, mark=*, mark size=1pt, mark options={solid}},
    generic/.style={black, evalline},
    noisy/.style={gray, evalline, densely dotted},
    log_pery/.style={dashed, LimeGreen, evalline},
    log_pery_log_psdn/.style={solid, OliveGreen, evalline},
    log_apost/.style={SkyBlue, evalline},
    log_aprior/.style={densely dotted, Blue, evalline},
    log_apost_log_aprior/.style={dashdotdotted, Magenta, evalline},
}

\pgfplotsset{%
    cocktail-amb-def/.style={Plum},
    aircraft-747-400-interior-midflight/.style={SkyBlue},
    factory1/.style={BrickRed},
    highway-traffic-heavy-traffic/.style={LimeGreen},
    mod_pink/.style={Lavender},
    mod_white/.style={Black},
    vacuum_cleaner/.style={Orange},
    myscatter/.style={only marks, mark options={draw=none}, mark size=0.15pt, mark=*},
}

\newenvironment{customlegend}[1][]{%
    \begingroup
    \csname pgfplots@init@cleared@structures\endcsname
    \pgfplotsset{#1}%
}{%
    \csname pgfplots@createlegend\endcsname
    \endgroup
}%

\def\addlegendimage{\csname pgfplots@addlegendimage\endcsname}

\pgfplotsset{%
    colormap={parula}{%
        rgb255=(53,42,135)
        rgb255=(15,92,221)
        rgb255=(18,125,216)
        rgb255=(7,156,207)
        rgb255=(21,177,180)
        rgb255=(89,189,140)
        rgb255=(165,190,107)
        rgb255=(225,185,82)
        rgb255=(252,206,46)
        rgb255=(249,251,14)}}

\tikzsetfigurename{Journal2018-figure}

\DeclareAcronym{ANOVA}{short=ANOVA, long={analysis of variance}}
\DeclareAcronym{AMS}{short=AMS, short-plural-form={amplitude modulation spectra}, long-plural-form={amplitude modulation spectra}, long={amplitude modulation spectrum}}
\DeclareAcronym{AR}{short=AR, long={auto-regressive}}
\DeclareAcronym{ASR}{short=ASR, long={automatic speech recognition}}
\DeclareAcronym{BECOCO}{short=BECOCO, long={phase-(b)lind (e)stimator of (co)mplex (co)efficients}}
\DeclareAcronym{cIRM}{short=cIRM, long={complex ideal ratio mask}}
\DeclareAcronym{CDF}{short=CDF, long={cumulative distribution function}}
\DeclareAcronym{CMVN}{short=CMVN, long={cepstral mean and variance normalization}}
\DeclareAcronym{CSNE}{short=CSNE, long={concatenated short noise excerpts}}
\DeclareAcronym{DFT}{short=DFT, long={discrete Fourier transform}}
\DeclareAcronym{DNN}{short=DNN, long={deep neural network}}
\DeclareAcronym{EM}{short=EM, long={expectation maximization}}
\DeclareAcronym{ESTOI}{short=ESTOI, long={extended \acl{STOI}}}
\DeclareAcronym{GAN}{short=GAN, long={generative adversarial network}}
\DeclareAcronym{GMM}{short=GMM, long={Gaussian mixture model}}
\DeclareAcronym{HMM}{short=HMM, long={hidden Markov model}}
\DeclareAcronym{IBM}{short=IBM, long={ideal binary mask}}
\DeclareAcronym{IDFT}{short=IDFT, long={inverse discrete Fourier transform}}
\DeclareAcronym{IRM}{short=IRM, long={ideal ratio mask}}
\DeclareAcronym{IS}{short=IS, long={Itakura-Saito}}
\DeclareAcronym{LSA}{short=LSA, long={log-spectral amplitude estimator}}
\DeclareAcronym{LSTM}{short=LSTM, long={long short-term memory}}
\DeclareAcronym{MAP}{short=MAP, long={maximum \emph{a posteriori}}}
\DeclareAcronym{MFCC}{short=MFCC, long={Mel-frequency cepstral coefficient}}
\DeclareAcronym{MLSE}{short=MLSE, long={machine-learning spectral envelope}}
\DeclareAcronym{ML}{short=ML, long={machine-learning}}
\DeclareAcronym{MMSE}{short=MMSE, long={minimum mean-squared error}}
\DeclareAcronym{MOSIE}{short=MOSIE, long={(M)MSE estimation with (o)ptimizable (s)peech (m)odel and (i)nhomogeneous (e)rror criterion}}
\DeclareAcronym{MSE}{short=MSE, long={mean-squared error}}
\DeclareAcronym{MUSHRA}{short=MUSHRA, long={multi-stimulus test with hidden reference and anchor}}
\DeclareAcronym{MoG}{short=MoG, long-plural-form={mixtures of Gaussians}, long=mixture of Gaussians}
\DeclareAcronym{NAT}{short=NAT, long={noise aware training}}
\DeclareAcronym{NMF}{short=NMF, long={non-negative matrix factorization}}
\DeclareAcronym{OMLSA}{short=OMLSA, long={optimally modified log-spectral amplitude estimator}}
\DeclareAcronym{PDF}{short=PDF, long={probability density function}}
\DeclareAcronym{PESQ}{short=PESQ, long={Perceptual Evaluation of Speech Quality}}
\DeclareAcronym{PLP}{short=PLP, long={Perceptive Linear Prediction}}
\DeclareAcronym{POLQA}{short=POLQA, long={perceptual objective listening quality analysis}}
\DeclareAcronym{WBPOLQA}{short=WB-POLQA, long={wideband perceptual objective listening quality analysis}}
\DeclareAcronym{PSD}{short=PSD, long={power spectral density}}
\DeclareAcronym{RASTA}{short=RASTA, long={relative spectral}}
\DeclareAcronym{ReLU}{short=ReLU, long={rectifying linear unit}}
\DeclareAcronym{SNR}{short=SNR, long={signal-to-noise ratio}}
\DeclareAcronym{SNRNAT}{short=SNR-NAT, long={\ac{SNR} based noise aware training}}
\DeclareAcronym{SPP}{short=SPP, long={speech presence probability}}
\DeclareAcronym{STFT}{short=STFT, long={short-time Fourier transform}}
\DeclareAcronym{ISTFT}{short=ISTFT, long={inverse short-time Fourier transform}}
\DeclareAcronym{VTS}{short=VTS, long-plural-form=vector Taylor series, short-plural-form=VTS, long=vector Taylor series}
\DeclareAcronym{STOI}{short=STOI, long={short-time objective intelligibility}}
\DeclareAcronym{STSA}{short=STSA, long={short-term spectral amplitude estimator}}
\DeclareAcronym{SegNR}{short=SegNR, long={segmental noise reduction}}
\DeclareAcronym{SegSNR}{short=SegSNR, long={segmental \acs{SNR}}}
\DeclareAcronym{SegSSNR}{short=SegSSNR, long={segmental speech \acs{SNR}}}
\DeclareAcronym{TCS}{short=TCS, long={temporal cepstrum smoothing}}
\DeclareAcronym{tSNE}{short=t-SNE, long={t-distributed stochastic neighbor embedding}}
\DeclareAcronym{VAD}{short=VAD, long={voice activity detector}}

\newcommand{\mysymbol}[5][]{%
    \expandafter\newcommand\csname #2\endcsname{#3}%
}

\newcommand{\mynotation}[5][]{%
    \expandafter\newcommand\csname #2\endcsname{#3}%
}

\newcommand{\sVec}[1]{\mathbf{#1}}
\newcommand{\sEst}[1]{\hat{#1}}
\newcommand{\sAvg}[1]{\overline{#1}}
\newcommand{\sMod}[1]{\tilde{#1}}

\newcommand{\sLog}[1]{{#1}^\text{(log)}}

\mysymbol[ak]{sFreqIdx}{k}{k}{frequency bin index}
\mysymbol[al]{sFrameIdx}{\ell}{\ell}{segment index}
\mysymbol[at]{sSampleIdx}{t}{t}{sample index}
\mysymbol[aq]{sCepsIdx}{q}{q}{cepstral index}
\newcommand{\sIdxPair}{{\sFreqIdx, \sFrameIdx}}

\newcommand{\sWin}{\omega}
\mysymbol[kwa]{sWinA}{\sWin_\text{a}}{\sWin_\text{a}(\cdot)}{spectral analysis window in the time domain}
\mysymbol[kws]{sWinS}{\sWin_\text{s}}{\sWin_\text{s}(\cdot)}{synthesis window in the time domain}

\mysymbol[cn]{sNoise}{n}{n_\sSampleIdx}{noise signal at sample $\sSampleIdx$}
\mysymbol[cs]{sSpeech}{s}{s_\sSampleIdx}{speech signal at sample $\sSampleIdx$}
\mysymbol[cy]{sNoisy}{y}{y_\sSampleIdx}{noisy signal at sample $\sSampleIdx$}
\newcommand{\sInput}{\sNoisy}

\mysymbol[cyi]{sFiltIn}{\sInput_\sFrameIdx}{\sInput_\sFrameIdx}{input of adaptive recursive smoothing filter at segment~$\sFrameIdx$}
\mysymbol[cyo]{sFiltOut}{\sAvg{\sInput}_\sFrameIdx}{\sAvg{\sInput}_\sFrameIdx}{output of adaptive recursive smoothing filter at segment~$\sFrameIdx$}
\newcommand{\sFiltOld}{\sAvg{\sInput}_{\sFrameIdx - 1}}

\mysymbol[ephi]{sPhase}{\Phi}{\Phi^x_{\sIdxPair}}{phase of the complex \acs{STFT} coefficients of signal component $x$}

\newcommand{\sNoiseF}{N}
\newcommand{\sNoiseFI}{\sNoiseF_{\sIdxPair}}
\newcommand{\sNoiseFIEst}{\sEst{\sNoiseF}_{\sIdxPair}}

\newcommand{\sSpeechF}{S}

\newcommand{\sSpeechFIEst}{\sEst{\sSpeechF}_{\sIdxPair}}

\newcommand{\sSpeechFI}{\sSpeechF_{\sIdxPair}}

\mysymbol[dA]{sSpeechMagF}{A}{\sSpeechMagFI}{magnitude of complex speech coefficients}
\newcommand{\sSpeechMagFI}{\sSpeechMagF_{\sIdxPair}}

\newcommand{\sNoisyF}{Y}
\newcommand{\sNoisyFI}{\sNoisyF_{\sIdxPair}}

\mysymbol[kfy]{sFunc}{\mathcal{F}}{\mathcal{F}(\cdot)}{arbitrary function of a random variable}

\mysymbol[hc]{sCorrection}{\mathcal{C}}{\mathcal{C}}{correction factor to compensate the bias of adaptive recursive smoothing filters via scaling (see Chapter~\ref{chp:AdaptiveSmoothing})}
\mysymbol[hc1]{sCorrectionMC}{\sCorrection_\text{MC}}{\sCorrection_\text{MC}}{correction factor obtained from Monte-Carlo simulations}

\mysymbol[hca]{sCorrectionAlt}{\sCorrection^\text{(a)}}{\sCorrection^\text{(a)}}{correction factor for adaptive recursive smoothing filter used for the alternative correction method (see Chapter~\ref{chp:AdaptiveSmoothingNonIterative})}
\mysymbol[fy]{sFix}{\sAvg{\sNoisy}^\text{(fix)}}{\sAvg{\sNoisy}^\text{(fix)}_\sIterationIdx}{fixed value used to replace $\sFiltOld$ in $\sSmoothAdapt$ which is updated iteratively to estimate the bias of adaptive recursive smoothing filter (see Section~\ref{sec:AdaptiveSmoothing:ExpectedValue})}
\mysymbol[hg]{sWienerCorr}{\mathcal{G}}{\mathcal{G}_\sIdxPair}{time-varying correction factor to compensate the bias of recursive adaptive smoothing factors using scaling (see Chapter~\ref{chp:AdaptiveSmoothing})}

\mysymbol[hga]{sCorrectionAltI}{\sWienerCorr^\text{(a)}_\sIdxPair}{\sWienerCorr^\text{(a)}_\sIdxPair}{time-varying correction factor to compensate the bias of adaptive recursive smoothing factors using the alternative correction method (see Chapter~\ref{chp:AdaptiveSmoothingNonIterative})}
\mysymbol[hcamc]{sCorrectionAltMC}{\sCorrection^\text{(a)}_\text{MC}}{\sCorrection^\text{(a)}_\text{MC}}{correction factor of the alternative correction method determined using Monte-Carlo simulations}

\mysymbol[fym]{sModFix}{\mathring{\sAvg{\sNoisy}}^\text{(fix)}}{\mathring{\sAvg{\sNoisy}}^\text{(fix)}_i}{fixed value used to replace $\sFiltOld$ in $\sSmoothAdapt$ for the alternative correction method in Section~\ref{sec:AdaptiveSmoothingNI:ParameterEstimation}}

\mysymbol[elx]{sVarF}{\Lambda}{\Lambda^x_\sIdxPair}{spectral \acs{PSD} of signal component $x$}
\mysymbol[elxgy]{sVarFXGY}{\Lambda}{\Lambda^{x|y}_\sIdxPair}{spectral \acs{PSD} of signal component $x$ given the value of $y$}

\newcommand{\sVarNoiseF}{\sVarF^\text{\sNoise}_\sIdxPair}
\newcommand{\sVarNoiseFn}{\sVarF^\text{\sNoise}}

\newcommand{\sVarSpeechF}{\sVarF^\text{\sSpeech}_\sIdxPair}

\newcommand{\sVarSpeechFEst}{\sEst{\sVarF}^\text{\sSpeech}_\sIdxPair}
\mysymbol[el1ml]{sVarSpeechFEstML}{\sEst{\sVarF}^{\sSpeech, \text{ml}}_\sIdxPair}{\sEst{\sVarF}^{\sSpeech, \text{ml}}_\sIdxPair}{maximum likelihood estimate of the speech \acs{PSD}}

\newcommand{\sVarNoiseFEst}{\sEst{\sVarF}^{\sNoise}_\sIdxPair}

\newcommand{\sVarNoiseFEstP}{\sEst{\sVarF}^{\sNoise}_{\sFreqIdx, \sFrameIdx - 1}}

\newcommand{\sVarSpeechCEst}{\sEst{\sVarF}^{\sSpeech}_{\sCepsIdx, \sFrameIdx}}
\newcommand{\sVarSpeechCEstP}{\sEst{\sVarF}^{\sSpeech}_{\sCepsIdx, \sFrameIdx - 1}}
\newcommand{\sVarSpeechCEstML}{\sEst{\sVarF}^{\sSpeech, \text{ml}}_{\sCepsIdx, \sFrameIdx}}

\newcommand{\sProb}{P}
\mysymbol[kfp]{sCDF}{F}{F(\cdot)}{cumulative density function}
\mysymbol[kf]{sPDF}{f}{f(\cdot)}{probability density function}

\mysymbol[kfmod]{sPDFModel}{\sMod{\sPDF}}{\sMod{\sPDF}(\cdot|\cdot)}{model distribution used in Section~\ref{sec:AdaptiveSmoothing:TransitionDensities} for estimating the bias of adaptive recursive smoothing filters}
\mysymbol[kg]{sPDFMarginal}{\sMod{g}}{\sMod{g}(\cdot|\cdot)}{marginalized distribution used in Section~\ref{sec:AdaptiveSmoothing:TransitionDensities} for estimating the bias of adaptive recursive smoothing filters}

\mysymbol[gslog]{sStdLN}{\lambda_\text{log($\mathcal{N}$)}}{\lambda_\text{log($\mathcal{N}$)}}{variance of the log-normal distribution}
\mysymbol[gmlog]{sMeanLN}{\mu_\text{log($\mathcal{N}$)}}{\mu_\text{log($\mathcal{N}$)}}{mean of the log-normal distribution}

\mysymbol[ga]{sSmoothConst}{\alpha}{\alpha}{smoothing constant of first-order recursive smoothing filters}

\mysymbol[kaa]{sSmoothAdapt}{\sSmoothConst(\sFiltIn, \sFiltOld)}{\sSmoothConst(\cdot, \cdot)}{adaptive smoothing factor}
\mysymbol[kzx]{sBiasFrac}{\mathcal{Z}}{\mathcal{Z}(x)}{bias of the recursive smoothing factor given a fixed value $x$ for the previous filter output}

\mynotation[statexpect]{sExpect}{\mathbb{E}}{\mathbb{E}\{\cdot\}}{expectation operator}

\mysymbol[kb]{sBCDistance}{\mathcal{B}}{\mathcal{B}(\cdot, \cdot)}{Bhattacharyya distance of two \acsp{PDF}}
\mysymbol[ke]{sBC}{\eta}{\eta(\cdot, \cdot)}{Bhattacharyya coefficient of two \acsp{PDF}}
\mysymbol[exi]{sPriorSNR}{\xi}{\xi_{\sIdxPair}}{\emph{a priori} \acs{SNR} $\sPriorSNRI = \sVarSpeechF / \sVarNoiseF$}
\mysymbol[gximl]{sPriorSNRMinML}{\xi^\text{ml}_\text{min}}{\xi^\text{ml}_\text{min}}{lower limit used in the maximum likelihood estimator of the \emph{a posteriori} \acs{SNR}}
\newcommand{\sPriorSNRI}{\sPriorSNR_{\sIdxPair}}

\mysymbol[eg]{sPostSNR}{\gamma}{\gamma_{\sIdxPair}}{\emph{a posteriori} \acs{SNR} $\sPostSNRI = |\sNoisyFI|^2 / \sVarNoiseF$}
\newcommand{\sPostSNRI}{\sPostSNR_{\sIdxPair}}

\mysymbol[ez]{sEphraimSubst}{\zeta}{\zeta_\sIdxPair}{equals Wiener gain times \emph{a posteriori} \acs{SNR} which often occurs in clean speech estimators}

\mysymbol[hh1]{sHypoSpeech}{\mathcal{H}_1}{\mathcal{H}_1}{speech presence hypothesis used in \acs{SPP}-based noise \acs{PSD} estimator}
\mysymbol[hh0]{sHypoNoise}{\mathcal{H}_0}{\mathcal{H}_0}{speech absence hypothesis used in \acs{SPP}-based noise \acs{PSD} estimator}
\mysymbol[gxih1]{sSNROpt}{\sPriorSNR_{\sHypoSpeech}}{\sPriorSNR_{\sHypoSpeech}}{fixed \emph{a priori} \acs{SNR} used in \acs{SPP}-based noise \acs{PSD} estimator}
\mysymbol[gasppf]{sSmoothSPPParam}{\sSmoothConst^\text{(fix)}_\text{SPP}}{\sSmoothConst^\text{(fix)}_\text{SPP}}{fixed smoothing constant used in \acs{SPP}-based noise \acs{PSD} estimator}
\mysymbol[kaspp]{sSmoothSPP}{\sSmoothConst_\text{SPP}(\sFiltIn, \sFiltOld)}{\sSmoothConst_\text{SPP}(\cdot, \cdot)}{adaptive smoothing factor of the \acs{SPP}-based noise \acs{PSD} estimator}
\newcommand{\sSmoothCI}{\sSmoothConst_{\sCepsIdx, \sFrameIdx}}
\mysymbol[gkv]{sBiasC}{\varkappa}{\varkappa}{bias correction term used in \acs{TCS}}

\mysymbol[kathr]{sSmoothAsymmetric}{\sSmoothConst_\text{Thr}(\sFiltIn, \sFiltOld)}{\sSmoothConst_\text{Thr}(\cdot, \cdot)}{adaptive smoothing factor of the threshold based noise \acs{PSD} estimator}
\mysymbol[gathra]{sSmoothUp}{\sSmoothConst^\uparrow}{\sSmoothConst^\uparrow}{smoothing factor of the threshold based noise \acs{PSD} estimator if the threshold is exceeded}
\mysymbol[gathrb]{sSmoothDown}{\sSmoothConst^\downarrow}{\sSmoothConst^\downarrow}{smoothing factor of the threshold based noise \acs{PSD} estimator if the threshold is not exceeded}

\mysymbol[gt]{sParameters}{\boldsymbol{\theta}}{\boldsymbol{\theta}}{vector of parameters}
\mysymbol[kphi]{sMiscFunction}{\phi}{\phi(\cdot)}{arbitrary function of the present and past inputs of a filter}
\mysymbol[ar]{sScale}{r}{r}{scaling factor}
\mysymbol[km]{sMeanFunction}{m}{m(\cdot)}{function that extracts mean of a \acs{PDF} from a set of parameters}
\mysymbol[go]{sOverlapFactor}{\Omega}{\Omega}{overlap of the \acs{STFT} segments}

\newcommand{\sIterationIdx}{i}

\mysymbol[bK]{sNumDFTBins}{K}{K}{number of \acs{DFT} coefficients}

\mysymbol[tLogErr]{sLogErr}{\text{LogErr}}{\text{LogErr}}{total log-error distortion}
\mysymbol[tLogErrOver]{sLogErrOver}{\text{LogErr}_\uparrow}{\text{LogErr}_\uparrow}{overestimation of the total log-error distortion}
\mysymbol[tLogErrPnder]{sLogErrUnder}{\text{LogErr}_\downarrow}{\text{LogErr}_\downarrow}{underestimation of the total log-error distortion}
\mysymbol[galogerr]{sSmoothLogErr}{\sSmoothConst^\text{(fix)}_\text{LogErr}}{\sSmoothConst_\text{LogErr}}{fixed smoothing constant to obtain reference noise \acs{PSD} for the log-error distortion measure}

\mysymbol[gadd]{sSmoothDD}{\sSmoothConst_\text{DD}}{\sSmoothConst_\text{DD}}{fixed smoothing constant of the decision-directed approach}

\mysymbol[gk]{sKurt}{\kappa}{\kappa_x}{Kurtosis of the signal $x$}
\mysymbol[gkd]{sLogKurtRatio}{\Delta\sLog{\sKurt}}{\Delta\sLog{\sKurt}}{log-kurtosis ratio}
\mysymbol[sn]{sSetNoise}{\mathbb{L}^\text{(\sNoise)}_\sFreqIdx}{\mathbb{L}^\text{(\sNoise)}_\sFreqIdx}{set of segments in a frequency band~$\sFreqIdx$ where noise is dominant}

\mysymbol[bL]{sNumSegments}{L}{L}{number of segments}

\mysymbol[kw]{sLambertW}{\mathcal{W}}{\mathcal{W}(\cdot)}{Lambert's $W$-function}

\newcommand{\sNoisyLog}{\sLog{\sNoisy}}
\newcommand{\sNoisyLogI}{\sNoisyLog_{\sIdxPair}}

\mysymbol[emx]{sMeanLog}{\mu}{\mu^{x}_{\sIdxPair}}{mean of the signal component $x$ in the log-spectral domain}

\mysymbol[elx]{sCovLog}{\lambda}{\lambda^x_{\sIdxPair}}{variance of the signal component $x$ in the log-spectral domain}

\mysymbol[gr]{sCorr}{\rho}{\rho}{correlation coefficient}

\mysymbol[kn]{sNormalDistr}{\mathcal{N}}{\mathcal{N}(x|\cdot, \cdot)}{normal distribution of $x$ with mean (first parameter) and variance (second parameter)}
\mysymbol[knc]{sNormalDistrC}{\sNormalDistr_{\mathbb{C}}}{\sNormalDistr_{\mathbb{C}}(x|\cdot, \cdot)}{complex circular-symmetric normal distribution of $x$ with mean (first parameter) and variance (second parameter)}
\mysymbol[gb]{sCompression}{\beta}{\beta}{parameter of exponential compression function $\sMMSEFun(S_\sIdxPair) = |S_\sIdxPair|^\sCompression$}

\mysymbol[gn]{sShape}{\nu}{\nu}{shape parameter of the $\chi$-distribution}

\mysymbol[dG]{sGain}{G}{G_\sIdxPair}{spectral gain}
\mysymbol[dGL]{sGainLimI}{\sMod{\sGain}_\sIdxPair}{\sMod{\sGain}_\sIdxPair}{limited spectral gain}
\mysymbol[fGMin]{sGainMin}{\sGain_{\text{min}}}{\sGain_{\text{min}}}{minimum value for spectral gain}
\newcommand{\sGainI}{\sGain_{\sIdxPair}}
\newcommand{\sGainIWiener}{\sGain^\text{Wiener}_{\sIdxPair}}
\newcommand{\sGainIIRM}{\sGain^\text{IRM}_{\sIdxPair}}
\newcommand{\sGainIIRMEst}{\sEst{\sGain}^\text{IRM}_{\sIdxPair}}

\mysymbol[kc]{sMMSEFun}{c}{c(\cdot)}{compression function used in \acs{MSE} optimal spectral clean speech estimators}

\newcommand{\sEuler}{\gamma}

\mysymbol[fq]{sState}{q}{q}{phoneme in Chapter~\ref{chp:SuperGaussian} and~\ref{chp:SuperGaussMixMax} or mixture of the speech \acs{GMM} in Chapter~\ref{chp:EnvelopeCombination}}
\mysymbol[bQ]{sNumStates}{Q}{Q}{number of phonemes in Chapter~\ref{chp:SuperGaussian} and~\ref{chp:SuperGaussMixMax} or number of mixtures in the speech \acs{GMM} in Chapter~\ref{chp:EnvelopeCombination}}

\mysymbol[dz]{sModel}{z}{z_\sIdxPair}{state indicator of the enhancement algorithm used for the combination in Chapter~\ref{chp:EnvelopeCombination}}

\mysymbol[fzmlse]{sModelVTS}{\sModel^\text{MLSE}}{\sModel^\text{MLSE}}{indicator of the \acs{MLSE} approach in Chapter~\ref{chp:EnvelopeCombination}}
\mysymbol[fzwf]{sModelWF}{\sModel^\text{WF}}{\sModel^\text{WF}}{indicator of the linear log-spectral filter in Chapter~\ref{chp:EnvelopeCombination}}
\mysymbol[fzlsa]{sModelLSA}{\sModel^\text{LSA}}{\sModel^\text{LSA}}{indicator of the \acs{LSA} in Chapter~\ref{chp:EnvelopeCombination}}

\mysymbol[db]{sSPP}{b}{b_\sIdxPair}{$\sSPPI = \sPDF_\sSpeech(\sNoisyLogI) \sCDF_\sNoise(\sNoisyLogI) / \sPDF_\sNoisy(\sNoisyLogI)$ used in the MixMax clean speech estimator}
\newcommand{\sSPPI}{\sSPP_{\sIdxPair}}

\mysymbol[kg]{sGamma}{\Gamma}{\Gamma(\cdot)}{gamma function}
\mysymbol[kmz]{sHypergeom}{\mathcal{M}}{\mathcal{M}(\cdot,\cdot;\cdot)}{confluent hypergeometric function}
\mysymbol[kpsi]{sDigamma}{\psi}{\psi(\cdot)}{Digamma function}
\mysymbol[kpsi1]{sTrigamma}{\psi_1}{\psi_1(\cdot)}{Trigamma function}

\mysymbol[cv]{sFeature}{v}{v_{i, \sFrameIdx}}{element of the feature vector $\sFeatureVec_\sFrameIdx$}

\newcommand{\sFeatureVec}{\sVec{\sFeature}}
\newcommand{\sFeatureVecY}{\sFeatureVec^{(\text{Per})}}
\newcommand{\sFeatureVecVarN}{\sFeatureVec^{(\text{PSD}_\text{\sNoise})}}
\newcommand{\sFeatureVecNAT}{\sFeatureVec^{(\text{NAT})}}
\newcommand{\sFeatureVecPrior}{\sFeatureVec^{(\text{prior})}}
\newcommand{\sFeatureVecPost}{\sFeatureVec^{(\text{post})}}
\newcommand{\sFeatureVecNorm}{\sFeatureVec^{(\text{SNR-NAT})}}
\newcommand{\sSFeatureVec}{\sMod{\sFeatureVec}}

\mysymbol[bHx]{sNumHiddenOutputs}{H}{H_x}{number of units in $x$th hidden layer}

\mysymbol[ge]{sBias}{\epsilon}{\epsilon}{bias in \acs{DNN} cost function}

\mysymbol[bV]{sFeatureDim}{V}{V}{dimensionality of the feature vector}

\mysymbol[nmfy]{sPerMat}{\sVec{Y}}{\sVec{Y}}{non-negative matrix of the noisy periodogram}

\mysymbol[nmfb]{sBaseMat}{\sVec{B}}{\sVec{B}}{non-negative matrix of the basis vectors}

\newcommand{\sActiEl}{w}

\mysymbol[nmfw]{sActiMat}{\sVec{W}}{\sVec{W}}{non-negative matrix of the activations}
\mysymbol[mw]{sActiVec}{\sVec{\sActiEl}}{\sVec{\sActiEl}_\sFrameIdx}{$\sFrameIdx$th column of the activation matrix $\sActiMat$}

\mysymbol[ss]{sFrameSet}{\mathbb{L}^\text{(\sSpeech)}}{\mathbb{L}^\text{(\sSpeech)}}{set of speech active segments}
\mysymbol[sq]{sFrameSetState}{\mathbb{L}^{(\sState)}}{\mathbb{L}^{(\sState)}}{set of segments belonging to a specific phoneme or mixture $\sState$}

\mysymbol[gd]{sSparsity}{\delta}{\delta}{controls sparsity of \acs{NMF} activations}

\mysymbol[bJ]{sCost}{J}{J}{cost function}

\mysymbol[tirm]{sIRM}{\text{IRM}}{\text{IRM}(\cdot)}{function that computes the ideal ratio mask from the noisy observation}

\mysymbol[bM]{sSegmentLength}{M}{M}{segment length in samples}
\mysymbol[bR]{sSegmentShift}{R}{R}{segment shift in samples}

\mysymbol[hw]{sMixingFun}{\mathcal{V}}{\mathcal{V}}{simplified mixing function used in \acs{VTS}-based \acs{MLSE} approach}
\mysymbol[hp]{sLinPoint}{\mathcal{P}_0}{\mathcal{P}_0}{linearization point for \acs{VTS} approach}

\newcommand{\sLearningRate}{\text{LR}}
\newcommand{\sEpoch}{E}

\newcommand{\sURL}{\url{https://www.inf.uni-hamburg.de/en/inst/ab/sp/publications/tasl2021-robust-se-rr.html}}

\input{colorconverter.tex}

\begin{document}

\title{SNR-Based Features and Diverse Training Data for Robust DNN-Based Speech Enhancement}

\author{Robert Rehr and~Timo Gerkmann}

\maketitle

\IEEEpeerreviewmaketitle

\begin{abstract}
    In this paper, we address the generalization of \ac{DNN} based speech enhancement to unseen noise conditions for the case that training data is limited in size and diversity.
    To gain more insights, we analyze the generalization with respect to (1) the size and diversity of the training data, (2) different network architectures, and (3) the chosen features. 
    To address (1), we train networks on the Hu noise corpus (limited size), the CHiME~3 noise corpus (limited diversity) and also propose a large and diverse dataset collected based on freely available sounds.
    To address (2), we compare a fully-connected feed-forward and a \ac{LSTM} architecture.
    To address (3), we compare three input features, namely logarithmized noisy periodograms, \ac{NAT} and the proposed \ac{SNRNAT}.
    We confirm that rich training data and improved network architectures help \acp{DNN} to generalize.
    Furthermore, we show via experimental results and an analysis using \ac{tSNE} that the proposed \ac{SNRNAT} features yield robust and level independent results in unseen noise even with simple network architectures and when trained on only small datasets, which is the key contribution of this paper.

\end{abstract}

\acresetall%

\section{Introduction}

Speech is used by humans for communication, e.g., to exchange ideas and emotions.
Speech is therefore an essential component of many applications that have emerged from increasingly powerful personal electronic devices such as hearing aids, mobile phones and virtual assistants.
Many devices are mobile and are therefore often used in noisy environments.
As a consequence, the microphones do not only capture the desired speech signal but also unwanted background noises.
Background noises are known to degrade the perceived quality of speech and are able to deteriorate the speech intelligibility.
To restore the quality and potentially also the intelligibility, speech enhancement algorithms are leveraged.
In this paper, we consider single-channel speech enhancement algorithms that allow the enhancement of noisy recordings obtained from a single microphone or the output of a spatial filter.

Single-channel speech enhancement has been a topic of research for many decades and various approaches have been introduced in the literature~\cite{boll_suppression_1979, ephraim_speech_1984, ephraim_bayesian_1992, schmidt_reduction_2008, mohammadiha_supervised_2013, lu_speech_2013, xu_regression_2015}.
Many approaches use a time-frequency representation, e.g., the \ac{STFT}, to enhance noisy input signals.
Often, the enhancement can be represented by a multiplication of the noisy Fourier coefficients with a real-valued gain function.
Conventional speech enhancement algorithms are often derived in a statistical framework.
The complex Fourier coefficients are modeled by parametric \acp{PDF} whose parameters are given by the speech \ac{PSD} and the noise \ac{PSD}.
The statistical framework then allows the derivation of statistically optimal estimators of the clean speech coefficients.
Depending on the statistical assumptions about speech and noise, different gain functions are obtained~\cite{ephraim_speech_1984, breithaupt_parameterized_2008, erkelens_tracking_2008, hendriks_log-spectral_2009}.
The unknown parameters of the speech \ac{PDF} and the noise \ac{PDF}, i.e., the speech \ac{PSD} and the noise \ac{PSD}, are estimated using algorithms based on statistical and signal processing models~\cite{ephraim_speech_1984, cohen_noise_2002, martin_noise_2001, breithaupt_novel_2008,gerkmann_unbiased_2012}.
Most of these algorithms are based on the assumption that the background noise changes more slowly than speech.
This makes conventional speech enhancement algorithms robust to many acoustic conditions, which is why they provide good results in moderately varying noise types, i.e., noises where the amplitude changes slowly in the \ac{STFT} bands.
This applies for example for passing cars on a busy street.
However, they lack the ability to track very fast changes of the background noise such as the cutlery in a restaurant or sudden speech bursts in a babble scene.
Consequently, transient noises are often not suppressed by conventional approaches.

The shortcomings of conventional speech enhancement algorithms motivated the use of \ac{ML} algorithms for speech enhancement.
Various \ac{ML} algorithms have been considered, e.g., codebooks~\cite{srinivasan_codebook_2006, rosenkranz_improving_2012, he_multiplicative_2017}, hidden Markov models and Gaussian mixture models~\cite{ephraim_bayesian_1992, sameti_hmm-based_1998, burshtein_speech_2002, zhao_hmm-based_2007, aroudi_hidden_2015} and \ac{NMF}~\cite{schmidt_reduction_2008, mohammadiha_supervised_2013, mohammadiha_transient_2014, simsekli_non-negative_2014}.
Today, \acp{DNN} have become a widely used tool for speech enhancement~\cite{maas_recurrent_2012, lu_speech_2013, weninger_deep_2014, xu_regression_2015, chazan_hybrid_2016, park_fully_2017, qian_speech_2017, zhao_convolutional-recurrent_2018}.
\Acp{DNN} potentially allow the approximation of any non-linear function on a limited range of the input space~\cite{hornik_approximation_1991}.
This flexibility allows \acp{DNN} to be used in various ways for speech enhancement.
In~\cite{suhadi_data-driven_2011, xia_wiener_2014, chinaev_optimal_2015, mirsamadi_causal_2016}, parts of conventional speech enhancement algorithms, e.g., the speech \ac{PSD} estimator or the noise \ac{PSD} estimator, have been replaced by \acp{DNN}.
Other approaches try to find a mapping from the noisy observation or features extracted from it to a masking function~\cite{wang_training_2014, weninger_discriminatively_2014, erdogan_phase-sensitive_2015}.
Suitable target functions and their effect on the enhancement performance have been analyzed in~\cite{wang_training_2014, erdogan_phase-sensitive_2015, wang_oracle_2016}.
Instead of using a masking function,  the clean speech coefficients have also been used as target of a \ac{DNN} in many approaches~\cite{lu_speech_2013, weninger_discriminatively_2014, xu_regression_2015, williamson_complex_2016}.
For this, various architectures of \acp{DNN} have been employed, e.g., feed-forward networks~\cite{lu_speech_2013, xu_regression_2015}, recurrent neural networks~\cite{maas_recurrent_2012} including long short-term memory cells~\cite{hochreiter_long_1997, erdogan_phase-sensitive_2015, weninger_speech_2015, chen_long_2017}, generative adversarial networks~\cite{pascual_segan_2017, michelsanti_conditional_2017}, convolutional neural networks~\cite{hui_convolutional_2015, park_fully_2017, zhao_convolutional-recurrent_2018} and WaveNet based approaches~\cite{van_den_oord_wavenet_2016, qian_speech_2017}.

Many studies show that \ac{DNN} based approaches yield higher performance than conventional speech enhancement approaches.
One of the concerns is however the generalization to unknown acoustic conditions~\cite{may_generalization_2014, xu_dynamic_2014, kumar_speech_2016, chazan_hybrid_2016, wang_joint_2017, kolbaek_speech_2017}.
On the one hand, the issue of generalization is encountered by leveraging large or diverse training data~\cite{xu_regression_2015, chen_large-scale_2016, kolbaek_speech_2017, pandey_cross-corpus_2020}.
On the other hand, estimations of the background noise have been included as inputs to a \ac{DNN} to improve the robustness in unseen acoustic conditions~\cite{xu_dynamic_2014, xu_regression_2015, kumar_speech_2016, wang_joint_2017}.
Inspired by \ac{NAT} proposed in~\cite{seltzer_investigation_2013}, a fixed noise estimate has been used in~\cite{xu_regression_2015} which is obtained from the first part of the noisy input signal.
The fixed estimate is replaced in dynamic \ac{NAT}~\cite{xu_dynamic_2014, kumar_speech_2016} by a time-varying noise estimate which is obtained from a conventional noise \ac{PSD} estimator.
Also \ac{DNN} based noise estimators have been considered to improve the robustness in unseen acoustic conditions~\cite{xu_dynamic_2014, mirsamadi_causal_2016}.
Further extensions of the approach in~\cite{xu_dynamic_2014} are presented in~\cite{wang_joint_2017}.
Here, additional input features such as the \ac{IRM}~\cite{wang_training_2014}, which are estimated using a \ac{DNN}, are used to increase the robustness.

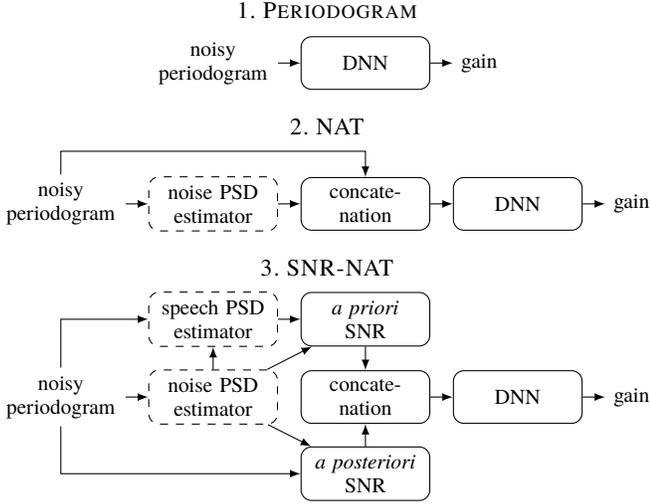
\begin{figure}[tb]
    \centering
    \small
    \textsc{1.\ Periodogram}\\[1.5ex]
    \begin{tikzpicture}[
        font={\footnotesize},
        node distance=0.3cm and 0.3cm,
        textblock/.style={align=center,
                          text width=1.5cm},
        conventional/.style={dashed},
        block/.style={rounded corners,
                      rectangle,
                      draw,
                      minimum width=1cm,
                      text width=1.5cm,
                      align=center,
                      minimum height=0.7cm},
    ]
    \node[textblock] (input) {noisy\\periodogram};
    \node[block, right=of input] (dnn) {DNN};
    \node[right=of dnn] (output) {gain};

    \draw[-latex] (input) -- (dnn);
    \draw[-latex] (dnn) -- (output);
\end{tikzpicture}\\[1.5ex]
    \textsc{2.\ \acs{NAT}}\\[1.5ex]
    \begin{tikzpicture}[
        font={\footnotesize},
        node distance=0.3cm and 0.3cm,
        textblock/.style={align=center,
                          text width=1.5cm},
        conventional/.style={dashed},
        block/.style={rounded corners,
                      rectangle,
                      draw,
                      minimum width=1cm,
                      text width=1.5cm,
                      align=center,
                      minimum height=0.7cm},
    ]
    \node[textblock] (input) {noisy periodogram};
    \node[block, conventional, right=of input] (noiseestimator) {noise PSD estimator};
    \node[block, right=of noiseestimator] (concat) {concatenation};
    \node[block, right=of concat] (dnn) {DNN};
    \node[right=of dnn] (output) {gain};

    \draw[-latex] (input) -- (noiseestimator);
    \draw[-latex] (input) -- ++(0, 0.7) -| (concat);
    \draw[-latex] (noiseestimator) -- (concat);
    \draw[-latex] (concat) -- (dnn);
    \draw[-latex] (dnn) -- (output);
\end{tikzpicture}\\[1.5ex]
    \textsc{3.\ \acs{SNRNAT}}\\[1.5ex]
    \begin{tikzpicture}[
        font={\footnotesize},
        node distance=0.3cm and 0.3cm,
        textblock/.style={align=center,
                          text width=1.5cm},
        conventional/.style={dashed},
        block/.style={rounded corners,
                      rectangle,
                      draw,
                      minimum width=1cm,
                      text width=1.5cm,
                      align=center,
                      minimum height=0.7cm},
    ]
    \node[textblock] (input) {noisy periodogram};
    \node[block, conventional, right=of input] (noiseestimator) {noise PSD estimator};
    \node[block, conventional, above=of noiseestimator] (speechestimator) {speech PSD estimator};
    \node[block, right=of noiseestimator] (concat) {concatenation};
    \node[block, right=of speechestimator] (snrpsd) {\emph{a priori} SNR};
    \node[block, below=of concat] (snrper) {\emph{a posteriori} SNR};
    \node[block, right=of concat] (dnn) {DNN};
    \node[right=of dnn] (output) {gain};

    \draw[-latex] (input) |- (speechestimator);
    \draw[-latex] (input) -- (noiseestimator);
    \draw[-latex] (input) |- (snrper);
    \draw[-latex] (snrper) -- (concat);
    \draw[-latex] (noiseestimator) -- (speechestimator);
    \draw[-latex] (noiseestimator) -- (snrpsd);
    \draw[-latex] (noiseestimator) -- (snrper);
    \draw[-latex] (speechestimator) -- (snrpsd);
    \draw[-latex] (snrpsd) -- (concat);
    \draw[-latex] (concat) -- (dnn);
    \draw[-latex] (dnn) -- (output);
\end{tikzpicture}\\[1.5ex]
    \caption{Block diagrams of the speech enhancement algorithms considered in this paper.
    Blocks with a dashed border indicate conventional algorithms.}%
    \label{fig:Introduction}
\end{figure}

In this paper, we focus on the generalization of \ac{DNN} based speech enhancement algorithms and consider the three approaches depicted in Fig.~\ref{fig:Introduction}.
All approaches predict a gain function, i.e., regression approaches as proposed in~\cite{xu_regression_2015} are not considered.
The input features are different though.
The approach depicted in the first row of Fig.~\ref{fig:Introduction} serves as a baseline and uses only the periodogram of the noisy input as feature.
The second approach is similar to dynamic \ac{NAT} presented in~\cite{xu_dynamic_2014, kumar_speech_2016}.
As shown in Fig.~\ref{fig:Introduction}, an estimate of the noise \ac{PSD} is appended to the noisy periodogram features.
Similar to~\cite{kumar_speech_2016}, we use the conventional noise \ac{PSD} estimator proposed in~\cite{gerkmann_unbiased_2012} to continuously update the noise \ac{PSD} estimate.
Last, we propose \ac{SNRNAT} features, which are related to the \emph{a priori} \ac{SNR} and the \emph{a posteriori} \ac{SNR} as defined in~\cite{ephraim_speech_1984}.
The corresponding \acp{SNR} are based on speech and noise \ac{PSD} estimates obtained from noisy speech using conventional enhancement approaches~\cite{gerkmann_unbiased_2012, breithaupt_novel_2008} as shown in the third row of Fig.~\ref{fig:Introduction}.
We show that noise datasets such as the Hu noise corpus~\cite{hu_corpus_2005} or the CHiME~3 noise corpus~\cite{barker_third_2015, barker_third_2017} are either not sufficiently large or diverse to train a general model using the standard logarithmized periodogram or \ac{NAT} features.
We propose a larger and more diverse training corpus using sounds from \url{freesound.org}.
It is used to confirm that increasing the size and the diversity of the noise data increases the robustness of \ac{DNN} based enhancement models.
The key contribution of this paper is that the generalization to unseen noise can also be addressed by using more informed features based on conventional signal processing approaches, i.e., using the proposed \ac{SNRNAT} features.
We show that using the \ac{SNRNAT} features, the training is more robust to insufficient training such that even a small and less diverse dataset allows the model to generalize to unseen acoustic conditions.
Further, we show that the \ac{DNN} based enhancement scheme becomes independent of the input's signal level if \ac{SNRNAT} features are employed.

For better understanding, we analyze and compare the statistics of the logarithmized periodogram, the \ac{NAT} features and the \ac{SNRNAT} features.
For this analysis, we use histograms and \ac{tSNE} for visualization~\cite{maaten_visualizing_2008}.
The results show that the \ac{SNRNAT} features are less dependent on changes of the overall level and the background noise.
Further, also the internal representation of the \ac{DNN} is less dependent on the background noise if \ac{SNRNAT} features are employed instead of \ac{NAT} features.

This paper extends our previous papers~\cite{rehr_robust_2018, rehr_analysis_2019} and provides for a more in depth analysis and evaluation.
It has the following structure.
In Section~\ref{sec:Conventional}, we recapitulate the conventional speech and noise \ac{PSD} estimators presented in~\cite{breithaupt_novel_2008, gerkmann_noise_2011, gerkmann_unbiased_2012} that are used to extract speech \ac{PSD} and noise \ac{PSD} as shown in the second and third row of Fig.~\ref{fig:Introduction}.
These estimates are later used in the \ac{NAT} and \ac{SNRNAT} features which are described in Section~\ref{sec:DNN}.
Furthermore, the \ac{DNN} based enhancement methods and the used network architectures are also presented in Section~\ref{sec:DNN}.
Section~\ref{sec:ExperimentalSetup} describes the evaluation and the training data which is used for the evaluation in Section~\ref{sec:InstrumentalEvaluation} and the analysis in Section~\ref{sec:Analysis}.
Section~\ref{sec:Conclusions} concludes this paper.

\section{Conventional Speech and Noise PSD Estimation}%
\label{sec:Conventional}

This section gives an overview of conventional speech enhancement in the \ac{STFT} domain.
The first part describes the steps required to enhance a signal in the \ac{STFT} domain and the following sections consider conventional speech and noise \ac{PSD} estimation algorithms.
Here, the conventional noise \ac{PSD} estimator which is based on~\cite{gerkmann_noise_2011, gerkmann_unbiased_2012} is recapitulated first.
Last, the speech \ac{PSD} estimator is considered which uses cepstral smoothing as described in~\cite{breithaupt_novel_2008}.
The speech and the noise \ac{PSD} estimates of these conventional algorithms also form the basis of the \ac{NAT} and \ac{SNRNAT} features used for the \ac{DNN}-based enhancement scheme discussed in Section~\ref{sec:DNN}.

\subsection{Speech enhancement in the STFT domain}%
\label{sec:STFTEnhancement}

This section describes how speech can be enhanced in the \ac{STFT} domain which is leveraged by various speech enhancement algorithms.
The presented procedure also applies to the \ac{DNN} based enhancement method presented in Section~\ref{sec:DNN}.
The \ac{STFT} of a time-domain signal is obtained by splitting it into overlapping segments and taking the Fourier transform of each segment after a tapered spectral analysis window has been applied.
This procedure results in the time-frequency representation of the clean speech signal $\sSpeechFI$, the noise signal $\sNoiseFI$ and the noisy input signal $\sNoisyFI$.
Here, $\sFreqIdx$ is the frequency index and $\sFrameIdx$ is the segment index.
In this work, a segment length of 32~ms is used and the segment shift is set to 16~ms, i.e., the segments overlap by 50~\%.
The considered input signals are sampled using a sampling rate of 8~kHz and consequently, the segment length corresponds to 256~samples and the segment shift to 128~samples.
Further, we use a square-root Hann window for spectral analysis.

The enhancement takes place in the spectral domain and can be expressed using a gain function $\sGainI$ which is applied to the noisy spectra.
Mathematically, the estimated clean speech coefficients are given by
\begin{equation}
    \label{eq:GainFunction}
    \sSpeechFIEst = \max(\sGainI \sNoisyFI, \sGainMin),
\end{equation}
where $\sGainMin$ is a lower limit of the gain function.
This lower limit has been found helpful to reduce artifacts and disturbances in the enhanced speech signal~\cite{berouti_enhancement_1979}.
In this work, we choose to set $\sGainMin$ to $-20~\text{dB}$.
A well known gain function is the Wiener filter
\begin{equation}
    \label{eq:WienerFilter}
    \sGainIWiener = \frac{\sVarSpeechF}{\sVarSpeechF + \sVarNoiseF},
\end{equation}
where $\sVarSpeechF$ and $\sVarNoiseF$ denote the speech \ac{PSD} and the noise \ac{PSD}, respectively.
The Wiener filter is the \ac{MMSE} optimal estimator of the clean speech coefficients if the speech coefficients~$\sSpeechFI$ and the noise coefficients~$\sNoiseFI$ are additive, uncorrelated and follow a complex circular-symmetric Gaussian distribution.
The former two assumptions are based on the physical properties of the interaction of multiple sound sources, i.e., speech and noise.
The latter assumption is often justified by the central limit theorem which can be applied due to the Fourier sum which needs to be evaluated for obtaining the spectral coefficients~\cite[Chapter~4]{brillinger_time_2001}.
The speech \ac{PSD}~$\sVarSpeechF$ and the noise \ac{PSD}~$\sVarNoiseF$ are estimated blindly from the noisy observation using~\cite{breithaupt_novel_2008, gerkmann_noise_2011, gerkmann_unbiased_2012}.
Both algorithms are summarized in the following sections.

After estimating the clean speech coefficients, the clean speech spectra $\sSpeechFIEst$ are transformed back to the time-domain.
A synthesis window is applied to the resulting time-domain segments where again a square-root Hann window is employed.
After that, the enhanced clean speech signal is obtained using an overlap-add procedure.

\subsection{Noise PSD Estimation}%
\label{sec:NonMLAlgorithms:NoisePSD}

In this work, we use the algorithm presented in~\cite{gerkmann_noise_2011, gerkmann_unbiased_2012} to estimate the noise \ac{PSD}.
This estimator allows the tracking of moderate changes in the background noise such as passing cars.
However, it cannot track transient disturbances like the cutlery noise in a restaurant.
In the remainder of this section, the algorithm is briefly introduced.

The noise \ac{PSD} estimator in~\cite{gerkmann_noise_2011, gerkmann_unbiased_2012} models the complex noisy coefficients under the hypotheses of speech presence~$\sHypoSpeech$ and speech absence~$\sHypoNoise$ using parametric distributions.
Given $\sHypoNoise$, the noisy observations equal $\sNoisyFI = \sNoiseFI$ while under $\sHypoSpeech$ the noisy coefficients are given by $\sNoisyFI = \sSpeechFI + \sNoiseFI$ which is a physically plausible assumption.
The speech coefficients~$\sSpeechFI$ and the noise coefficients~$\sNoiseFI$ are assumed to follow a complex circular-symmetric Gaussian distribution.
Accordingly, the likelihoods under the hypotheses $\sHypoNoise$ and $\sHypoSpeech$, i.e.,~$\sPDF(\sNoisyFI|\sHypoNoise)$ and $\sPDF(\sNoisyFI|\sHypoSpeech)$, are also modeled using Gaussian distributions. 
The \ac{SPP} is defined as the posterior probability $\sProb(\sHypoSpeech|\sNoisyFI)$ which can be obtained using Bayes' theorem.
The posterior used in~\cite{gerkmann_noise_2011, gerkmann_unbiased_2012}
\begin{equation}
    \label{eq:PosteriorDistribution}
    \sProb(\sHypoSpeech | \sNoisyFI) = {\left(1 + (1 + \sSNROpt) \exp\left(-\frac{|\sNoisyFI|^2}{\sVarNoiseFEstP} \frac{\sSNROpt}{1 + \sSNROpt}\right) \right)}^{-1},
\end{equation}
has been derived under the assumption that the prior $\sProb(\sHypoSpeech) = \sProb(\sHypoNoise) = 1/2$.
Here, a fixed \ac{SNR}~$\sSNROpt$ is used which is interpreted as the local \ac{SNR} that is expected if the hypothesis $\sHypoSpeech$ holds~\cite{gerkmann_noise_2011, gerkmann_unbiased_2012}.
The likelihood models~$\sPDF(\sNoisyFI | \sHypoNoise)$ and~$\sPDF(\sNoisyFI | \sHypoSpeech)$ have been used to formulate a speech detection problem in~\cite{gerkmann_unbiased_2012}.
By minimizing the total risk of error~\cite{gerkmann_unbiased_2012}, the optimal value $\sSNROpt = -15~\text{dB}$ has been found.

The posterior probability $\sProb(\sHypoSpeech|\sNoisyFI)$ is used to estimate the noise periodogram as
\begin{equation}
    \label{eq:MMSENoisePeriodogram}
    |\sNoiseFIEst|^2 = (1 - \sProb(\sHypoSpeech|\sNoisyFI)) |\sNoisyFI|^2 + \sProb(\sHypoSpeech|\sNoisyFI) \sVarNoiseFEstP.
\end{equation}
The estimated noise periodogram $|\sNoiseFIEst|^2$ is smoothed temporally to obtain an estimate of the noise \ac{PSD} as
\begin{equation}
    \label{eq:SPPSmooth}
    \sVarNoiseFEst = (1 - \sSmoothSPPParam) |\sNoiseFIEst|^2 + \sSmoothSPPParam \sVarNoiseFEstP,
\end{equation}
where $\sSmoothSPPParam$ is a fixed smoothing constant.
This estimator can be implemented in a speech enhancement framework by evaluating~\eqref{eq:PosteriorDistribution},~\eqref{eq:MMSENoisePeriodogram} and~\eqref{eq:SPPSmooth} for each frequency band~$\sFreqIdx$ when a new segment $\sFrameIdx$ is processed.
If the noise \ac{PSD} is strongly underestimated, the \ac{SPP} in~\eqref{eq:PosteriorDistribution} is overestimated, i.e., it is close to 1.
As a result, the noise periodogram in~\eqref{eq:MMSENoisePeriodogram} may no longer be updated.
To avoid such stagnations, the \ac{SPP} is set to a lower value if it has been stuck at 1 for a longer period of time~\cite{gerkmann_noise_2011, gerkmann_unbiased_2012}.

\subsection{Speech PSD Estimation}%
\label{sec:NonMLAlgorithms:SpeechPSD}

For estimating the speech \ac{PSD}~$\sVarSpeechF$, the \ac{TCS} approach described in~\cite{breithaupt_novel_2008} is employed.
In contrast to the commonly used decision-directed approach~\cite{ephraim_speech_1984}, this approach causes less isolated estimation errors, which may be perceived as annoying musical tones.
In this section, we recapitulate the main concepts of this algorithm.

Under the assumption that the spectral speech and noise coefficients follow a complex circular-symmetric Gaussian distribution, the limited maximum likelihood estimator is given by~\cite{ephraim_speech_1984}
\begin{equation}
    \label{eq:MaximumLikelihood}
    \sVarSpeechFEstML = \sVarNoiseFEst \max\left(\frac{|\sNoisyFI|^2}{\sVarNoiseFEst} - 1, \sPriorSNRMinML\right),
\end{equation}
where the $\max(\cdot)$ operator in combination with~$\sPriorSNRMinML$ is used to avoid negative speech \acp{PSD} and numerical issues in the following steps.
For the practical applicability, $\sVarNoiseF$ has been replaced by its estimate $\sVarNoiseFEst$.

The maximum likelihood estimate is transformed to the cepstral domain via
\begin{equation}
    \sVarSpeechCEstML = \text{IDFT}\{\log(\sVarSpeechFEstML)\},
\end{equation}
where $\sCepsIdx$ is the quefrency index and $\text{IDFT}(\cdot)$ denotes the inverse discrete Fourier transform.
In the cepstral domain, speech can be represented by using only a few coefficients: The speech spectral envelope, which reflects the impact of the vocal tract filter, is represented by the lower coefficients with $\sCepsIdx < 2.5~\text{ms}$ whereas the speech spectral fine structure, i.e., the fundamental frequency and its harmonics, is approximated by a single peak among the high cepstral coefficients.
This peak is also referred to as pitch peak.
The compact representation of speech is exploited by the \ac{TCS} approach by using a quefrency and time dependent smoothing factor $\sSmoothCI$ to smooth $\sVarSpeechCEstML$ as
\begin{equation}
    \sVarSpeechCEst = (1 - \sSmoothCI) \sVarSpeechCEstML + \sSmoothCI \sVarSpeechCEstP.
\end{equation}
For the cepstral coefficients that are associated with speech only little smoothing is applied while the remaining cepstral coefficients are strongly smoothed.
Accordingly, $\sSmoothCI$ is set close to 0 for the lower cepstral coefficients and close to 1 for the high coefficients.
In voiced segments, the $\sSmoothCI$ in close vicinity to the cepstral pitch peak are changed to values close to 0.

The cepstrally smoothed speech \ac{PSD}~$\sVarSpeechCEst$ is transformed back to the spectral domain as
\begin{equation}
    \sVarSpeechFEst = \exp\left(\text{DFT}\{\sVarSpeechCEst\} + \frac{\sEuler}{2} \right).
\end{equation}
As the smoothing in the cepstral domain results in a biased estimate~\cite{gerkmann_statistics_2009}, the correction term $\sEuler/2$ is added where $\sEuler \approx 0.5772\dots$ is the Euler constant.
In~\cite{breithaupt_novel_2008}, it has been argued that the bias of computing the expected value of a spectral quantity following a Gaussian distribution in the logarithmic domain amounts to the Euler constant.
Due to the smoothing, the estimate in the cepstral domain is between an instantaneous value and the expected value.
A more rigorous analysis of the bias is given in~\cite{gerkmann_statistics_2009}.

\section{DNN Based Speech Enhancement}%
\label{sec:DNN}

In this section, the \ac{DNN} based speech enhancement algorithms used in this paper are presented.
The first part of this section, considers the two \ac{DNN} architectures analyzed in this paper.
We consider a feed-forward network and an \ac{LSTM} based network~\cite{hochreiter_long_1997}.
After that, the input features are considered, which are used for both networks.

\subsection{Network Architectures}%
\label{sec:DNN:Architecture}

Similar to the conventional speech enhancement scheme described in Section~\ref{sec:STFTEnhancement}, the \ac{DNN} based enhancement schemes also operate in the \ac{STFT} domain.
In this paper, the \ac{DNN}'s task is to estimate an \ac{IRM}~\cite{wang_training_2014} using the features $\sFeature_{\sFreqIdx, \sFrameIdx}$ extracted from the noisy input signal.
The \ac{IRM} has been proposed in~\cite{wang_training_2014} and depends on the speech periodogram $|\sSpeechFI|^2$ and the noise periodogram $|\sNoiseFI|^2$.
It can be defined as
\begin{equation}
    \label{eq:IRM}
    \sGainIIRM = \frac{|\sSpeechFI|^2}{|\sSpeechFI|^2 + |\sNoiseFI|^2}.
\end{equation}
While during training the actual speech and noise periodogram are available to compute the \ac{IRM}, a trained \ac{DNN} is used to predict the \ac{IRM} during processing.
The predicted \ac{IRM} is used to estimate the clean speech coefficients~$\sSpeechFIEst$ as in~\eqref{eq:GainFunction}.
Other targets such as ideal binary masks~\cite{wang_training_2014} or complex ideal ratio masks~\cite{williamson_complex_2016} are not considered and could potentially exhibit a different behavior than the employed \ac{IRM} target.

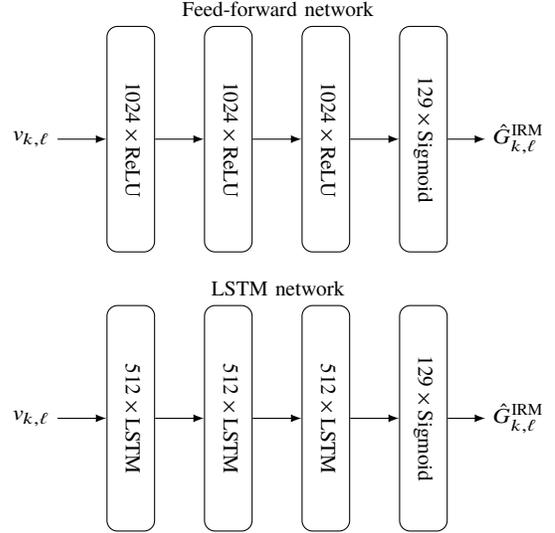
\begin{figure}[tb]
    \centering%
    \footnotesize{Feed-forward network}\\[1ex]
    \begin{tikzpicture}[%
    block/.style={rectangle,
                  draw,
                  rounded corners,
                  minimum width=5ex,
                  minimum height=3cm},
    ]

    \node (input) {$\sFeature_{\sFreqIdx, \sFrameIdx}$};
    \node[block, right=0.65cm of input]  (layer1) {\rotatebox{-90}{$1024 \times \text{ReLU\strut}$}};
    \node[block, right=0.65cm of layer1] (layer2) {\rotatebox{-90}{$1024 \times \text{ReLU\strut}$}};
    \node[block, right=0.65cm of layer2] (layer3) {\rotatebox{-90}{$1024 \times \text{ReLU\strut}$}};
    \node[block, right=0.65cm of layer3] (outputlayer) {\rotatebox{-90}{$129 \times \text{Sigmoid}$}};
    \node[right=0.5cm of outputlayer] (output) {$\sGainIIRMEst$};

    \draw[-latex] (input) -- (layer1);
    \draw[-latex] (layer1) -- (layer2);
    \draw[-latex] (layer2) -- (layer3);
    \draw[-latex] (layer3) -- (outputlayer);
    \draw[-latex] (outputlayer) -- (output);
\end{tikzpicture}\\[2ex]
    \footnotesize{\acs{LSTM} network}\\[1ex]
    \begin{tikzpicture}[%
    block/.style={rectangle,
                  draw,
                  rounded corners,
                  minimum width=5ex,
                  minimum height=3cm},
    ]

    \node (input) {$\sFeature_{\sFreqIdx, \sFrameIdx}$};
    \node[block, right=0.65cm of input]  (layer1) {\rotatebox{-90}{$512 \times \text{LSTM\strut}$}};
    \node[block, right=0.65cm of layer1] (layer2) {\rotatebox{-90}{$512 \times \text{LSTM\strut}$}};
    \node[block, right=0.65cm of layer2] (layer3) {\rotatebox{-90}{$512 \times \text{LSTM\strut}$}};
    \node[block, right=0.65cm of layer3] (outputlayer) {\rotatebox{-90}{$129 \times \text{Sigmoid}$}};
    \node[right=0.5cm of outputlayer] (output) {$\sGainIIRMEst$};

    \draw[-latex] (input) -- (layer1);
    \draw[-latex] (layer1) -- (layer2);
    \draw[-latex] (layer2) -- (layer3);
    \draw[-latex] (layer3) -- (outputlayer);
    \draw[-latex] (outputlayer) -- (output);
\end{tikzpicture}\\
    \caption{Block diagram of the \acs{DNN} architectures employed in this study.
    The upper diagram shows the feed-forward network, while the lower diagram shows the \acs{LSTM} network.}\label{fig:DNN:Architectures}
\end{figure}

The \ac{IRM} is estimated using two different network architectures.
Both networks differ in size and style to make it possible to provide an analysis on how the proposed features perform for different neural networks.
The first architecture has a feed-forward structure.
The input features are passed through three fully-connected hidden layers where each layer comprises 1024 \acp{ReLU}~\cite{glorot_deep_2011} as shown in the upper diagram Fig.~\ref{fig:DNN:Architectures}.
The last layer contains 129 units to match the dimensionality of the \ac{STFT} spectra after removing the mirror spectrum.
Sigmoid non-linearities are used to enforce the output to be in the same range as the \ac{IRM}, i.e., between zero and one.

The second architecture uses recurrent layers and is based on \ac{LSTM} cells~\cite{hochreiter_long_1997}.
Here, the input features are passed through three layers comprising 512 \ac{LSTM} cells.
Similar to the feed-forward network, the final layer is a fully connected layer with 129 sigmoid units.
The architecture is shown in the lower diagram of Fig.~\ref{fig:DNN:Architectures}.

The parameter choice has been inspired by other networks used in the literature, e.g.,~\cite{xu_regression_2015,erdogan_phase-sensitive_2015,chen_long_2017,sun_multiple-target_2017}, and resembles loosely the structures of those architectures.

\subsection{Features}%
\label{sec:Features}

In this section, the input features that are used in combination with both \ac{DNN} architectures are considered.

First, the logarithmized noisy periodogram features are tackled.
This feature forms the basis of the \ac{NAT} features, which are described afterwards.
The feature vector for the logarithmized periodogram of the noisy input coefficients has also been employed in various other studies~\cite{xu_dynamic_2014, weninger_discriminatively_2014, xu_regression_2015, mirsamadi_causal_2016}.
It is defined as
\begin{equation}
    \sFeatureVecY_\sFrameIdx = {[\log(|\sNoisyF_{1, \sFrameIdx}|^2), \ldots, \log(|\sNoisyF_{\sNumDFTBins, \sFrameIdx}|^2)]}^T.
\end{equation}
The vector transpose is denoted by $\cdot^T$ and $\sNumDFTBins$ is the number sampling points of the discrete Fourier transform after omitting the mirror spectrum, i.e., $\sNumDFTBins = 129$.
Please note that sometimes also (linear) transforms of logarithmized periodograms are used~\cite{wang_exploring_2013, wang_multiobjective_2018}, e.g., to obtain cepstral representations~\cite{davis_comparison_1980, shao_incorporating_2007}, which may change results slightly.
However, in this study we focus on the presented compact features for conciseness.

Given only the log-spectral coefficients of the noisy observation, the \ac{DNN} learns from the training data to distinguish between the desired speech signal and the background noise.
As this is a potentially challenging task, \ac{NAT} has been proposed to improve the robustness of \ac{DNN} based speech enhancement in unseen acoustic conditions~\cite{seltzer_investigation_2013, xu_dynamic_2014, kumar_speech_2016, wang_joint_2017}.
For this, \ac{NAT} appends a logarithmized noise \ac{PSD} estimate to the logarithmized noisy periodogram.
The feature vector of the logarithmized noisy \ac{PSD} is given by
\begin{equation}
    \sFeatureVecVarN_\sFrameIdx = {[\log(\sVarNoiseFn_{1, \sFrameIdx}), \ldots, \log(\sVarNoiseFn_{\sNumDFTBins, \sFrameIdx})]}^T.
\end{equation}
The \ac{NAT} features are then given by the concatenation of $\sFeatureVecY$ and $\sFeatureVecVarN$ for each frame $\sFrameIdx$ as
\begin{equation}
    \sFeatureVecNAT_\sFrameIdx = {\left[{(\sFeatureVecY_\sFrameIdx)}^T, {(\sFeatureVecVarN_\sFrameIdx)}^T\right]}^T.
\end{equation}
Using \ac{NAT} features, i.e., appending the noise \ac{PSD} to the noisy log-spectra, doubles the dimensionality of the input features which results in a 258-dimensional vector.
In our experiments, the noise \ac{PSD} is estimated using the approach described in Section~\ref{sec:NonMLAlgorithms:NoisePSD}, i.e., based on~\cite{gerkmann_noise_2011, gerkmann_unbiased_2012}.
The noise \ac{PSD} is estimated from noisy data both during training and testing such that the network learns the estimation characteristics of the employed noise \ac{PSD} estimator.

In contrast to the \ac{NAT} features, where the noise \ac{PSD} is appended to the noisy input periodogram, we proposed to use the noise \ac{PSD} for normalization to obtain so-called \ac{SNRNAT} features~\cite{rehr_robust_2018, rehr_analysis_2019}.
The \ac{SNRNAT} features correspond to logarithmized versions of the \emph{a priori} \ac{SNR} $\sPriorSNRI = \sVarSpeechF / \sVarNoiseF$ and the \emph{a posteriori} \ac{SNR} $\sPostSNRI = |\sNoisyFI|^2 / \sVarNoiseF$.
Both features can also be stacked in feature vectors as
\begin{align}
    \label{eq:DNN:FeatureAPrior}
    \sFeatureVecPrior_\sFrameIdx &= {[\log(\sPriorSNR_{1, \sFrameIdx}), \ldots, \log(\sPriorSNR_{\sNumDFTBins, \sFrameIdx})]}^T,\\
    \label{eq:DNN:FeatureAPost}
    \sFeatureVecPost_\sFrameIdx &= {[\log(\sPostSNR_{1, \sFrameIdx}), \ldots, \log(\sPostSNR_{\sNumDFTBins, \sFrameIdx})]}^T.
\end{align}
Further, also the combination of both \ac{SNR} based features is considered which yields the \ac{SNRNAT} features
\begin{equation}
    \label{eq:DNN:SNRNAT}
    \sFeatureVecNorm_\sFrameIdx = {\left[{(\sFeatureVecPrior_\sFrameIdx)}^T, {(\sFeatureVecPost_\sFrameIdx)}^T\right]}^T
\end{equation}
The noise \ac{PSD}~$\sVarNoiseF$ is estimated again using the approach from Section~\ref{sec:NonMLAlgorithms:NoisePSD}~\cite{gerkmann_noise_2011, gerkmann_unbiased_2012}.
The speech \ac{PSD} is estimated as described in~\cite{breithaupt_novel_2008} and Section~\ref{sec:NonMLAlgorithms:SpeechPSD}.
For similar reasons to the \ac{NAT} features, the speech and the noise \ac{PSD} are estimated from noisy data both during training and testing.
The \emph{a priori} \ac{SNR} and the \emph{a posteriori} \ac{SNR} have been previously used in data-driven speech enhancement approaches~\cite{erkelens_general_2006, erkelens_data-driven_2007, fingscheidt_data-driven_2006, fingscheidt_environment-optimized_2008}, but these approaches did not use \acp{DNN}.
More interestingly, we show in~\cite{rehr_robust_2018, rehr_analysis_2019} that the \ac{SNRNAT} features result in more robust \ac{DNN} based speech enhancement algorithms especially if the size or the diversity of the training data is limited.

For the feed-forward architecture described in Section~\ref{sec:DNN:Architecture}, a temporal context is added to all input features.
For this, a super-vector $\sSFeatureVec_\sFrameIdx$ is created which stacks the features of the current frame and the features from three previous frames
\begin{equation}
    \sSFeatureVec_\sFrameIdx = {\left[\sFeatureVec_\sFrameIdx^T, \ldots, \sFeatureVec_{\sFrameIdx - 3}^T \right]}^T.
\end{equation}
This increases the dimensionality of the input features by a factor four, i.e., the dimensionality is raised from 129 to 516 (logarithmized periodogram) or from 258 to 1032 (\ac{NAT} and \ac{SNRNAT}), respectively.
The resulting feed-forward network has 2.7~million weights, if the log periodogram features are used, and 3.3~million weights, if \ac{NAT} or \ac{SNRNAT} is used.

Due to their recurrent structure, \ac{LSTM} networks are able to include context on their own and therefore, no additional context is used for this network type.
Hence, the input dimensionality for the \ac{LSTM} networks remains 129 (logarithmized perdiogram) or 258 (\ac{NAT} and \ac{SNRNAT}),  respectively.
The \ac{LSTM} network has 5.6 million weights, if the log periodogram features are used, and 5.8 million weights, if \ac{NAT} or \ac{SNRNAT} features are used.

\section{Experimental Setup}%
\label{sec:ExperimentalSetup}

The \ac{DNN} based speech enhancement algorithm described in Section~\ref{sec:DNN} is trained using three noise corpora: the Hu noise corpus~\cite{hu_corpus_2005} with the extension presented in~\cite{xu_multi-objective_2015}, the CHiME 3 noise corpus~\cite{barker_third_2015, barker_third_2017} and a custom noise set which has been created from sound packs available from the \url{freesound.org} website.
We refer to the Hu corpus and its extension just as Hu corpus to keep the naming scheme simple.
The noise sets vary in the amount of audio data and the variety of the noise material as described below.

The Hu noise corpus~\cite{hu_corpus_2005} comprises 100~non-speech sounds and the extension~\cite{xu_multi-objective_2015} adds another 15~sounds.
The sounds cover many different acoustic environments, but the recordings are generally short and their duration often does not exceed ten seconds.
The total duration of the sound material included in this corpus is about 14~minutes.

The background noises included in the CHiME~3 challenge~\cite{barker_third_2015, barker_third_2017} comprise four different acoustic environments: a ride on a bus, the interior of a cafe, a pedestrian area and a street junction.
These scenarios have been captured using a tablet equipped with six microphones.
As we consider single-channel speech enhancement algorithms in this work, only the recordings of the first microphone are employed.
Due to the low number of acoustic environments included in the CHiME~3 noise corpus, also the diversity of the noise corpus is rather limited.
However, the duration of the recordings is long and the total amount of noise data in this corpus amount to about 8.5~hours.

\begin{table}[tb]
    \centering
    \caption{Sound packs that form the large and diverse \protect\url{freesound.org} noise dataset. The packs can be downloaded by replacing <username> and <id> in \protect\url{freesound.org/people/<username>/packs/<id>} by the data below.}\label{tab:freesoundpacks}
    \begin{tabular}{m{2cm}p{\linewidth-3.5cm}}
        \toprule
        username & list of ids\\
        \midrule
        Robinhood76 & {\small 3238, 3246, 3667, 3668, 3729, 3830, 3870, 3873, 3971, 3979, 3980, 4024, 4025, 4026, 4036, 4058, 4065, 4149, 4364, 5589}\\
        rutgermuller & {\small 20158}\\
        \bottomrule
    \end{tabular}
\end{table}

The last dataset is constructed from freely available sounds taken from the \url{freesound.org} website.
The links to the sound packs used to create this data set are given in Table~\ref{tab:freesoundpacks}.
From the sound recordings in these packs, we discard all sounds whose duration is less than 30~seconds.
This results in 282~sound files from many different acoustic environments with a total duration of about 13.5~hours.
In contrast to the Hu noise corpus~\cite{hu_corpus_2005, xu_multi-objective_2015}, which is limited in size, and the CHiME~3 corpus, which is limited in diversity, this dataset exhibits a large amount of data and a high diversity.

The training data is generated by artificially corrupting sentences from the TIMIT training corpus~\cite{garofolo_timit_1993} by the noise data given in the datasets above.
For training, 3992~sentences from 462~speakers are used where the number of sentences spoken by male and female speakers is the same.
Further, we use only the compact and diverse sentences to ensure phonetic diversity.
The duration of the clean speech data is about two hours.
From these sentences, a training set of 100~hours is generated for each of the noise datasets given above.
For this, the sentences from TIMIT data set are allowed to be reused 49 times.
Each sentence, also the reused ones, is randomly embedded in one of the background noises of the respective training data set.
Furthermore, also the excerpt of the noise is randomly chosen for each sentence.
To allow the noise \ac{PSD} estimator described in Section~\ref{sec:NonMLAlgorithms:NoisePSD}~\cite{gerkmann_noise_2011, gerkmann_unbiased_2012} to adapt to the background noise, a two second long initialization period is added in the beginning of the noisy sentence.
After the feature extraction, the initialization period is removed from the training data.
Using the first two seconds for initialization is not an issue in many speech communication applications like hearing aids or telecommunications.
In contrast, the training would be strongly impacted by initialization artifacts, if the initialization period is not excluded from the training data.
As a consequence, the training data would not be representative for real-world data.
To make the \ac{DNN} aware of different levels of corruption, the input \ac{SNR} of each sentence is varied between $-10~\text{dB}$ and $15~\text{dB}$.
Additionally, also level variations are included in the training data to allow the \ac{DNN} to adapt to different overall levels of the input signal.
For this, the time-domain peak level of the clean speech signal is varied between $-26~\text{dB}$ and $-3~\text{dB}$ before being corrupted by the background noise which also affects the overall level of the noisy signal.
It is ensured that about $10~\%$ of the training data contain only noise to enable the \ac{DNN} to reject noise only regions.

The data described above is split into a training set and a validation set.
A portion of $15~\%$ of the data is used as a validation set and the remaining data is used for training.
To initialize the parameters of the respective \acp{DNN}, the Glorot method described in~\cite{glorot_understanding_2010} is used.
After that, the parameters are trained by minimizing the squared-error between the true and the predicted \ac{IRM} using stochastic gradient descent.
The error function is given by the squared error
\begin{equation}
    \sCost = \sum_{\sFreqIdx} \sum_{\sFrameIdx} \left|\sGainIIRMEst - \sGainIIRM\right|^2.
\end{equation}
The learning rate is reduced from 0.4 to 0.1 over the training epochs using an exponential decay as $\sLearningRate = \max(0.4 \cdot 0.95^{\sEpoch -1}, 0.1)$.
The symbols $\sLearningRate$ and $\sEpoch$ denote the learning rate and the current epoch, respectively.
The feed-forward networks have been trained for a maximum of 50 epochs while for the \ac{LSTM} networks the maximum number of epochs was set to 20.
Only the model with the lowest validation error is used for testing which is similar to using an early stopping strategy.
All networks have been trained using Keras~2.2.4 and Tensorflow~1.13.1.

\section{Instrumental Evaluation}%
\label{sec:InstrumentalEvaluation}

In this section, we analyze how the input features affect the performance of the \ac{DNN} based speech enhancement algorithms using instrumental measures.
For this, we employ \ac{ESTOI}~\cite{jensen_algorithm_2016}, \ac{POLQA}~\cite{noauthor_p.863_2018}, as well as, \ac{SegSSNR} and \ac{SegNR} described in~\cite{lotter_speech_2005}.
\Ac{ESTOI} is an instrumental measure of the speech quality, while \ac{POLQA} is a measure of the speech quality.
Generally, the difference between the enhanced signal and the noisy signal, i.e., the improvement over the noisy signal, is shown for \ac{ESTOI} and \ac{POLQA} to simplify the comparison of the algorithms.
\Ac{SegSSNR} and \ac{SegNR} are used to measure the amount of speech distortion and noise reduction of the respective processing scheme.

For testing, we use noises taken from NOISEX-92 database~\cite{varga_assessment_1993} and \url{freesound.org}.
We ensure that none of these noise types have been used during training, i.e., all noise types used for testing are unseen in terms of the noise realization.
Depending on the used training data also some of the realizations can be different.
From the NOISEX-92 database, we include the “factory 1”, “f16” and “destroyerops” environment.
Additionally, amplitude modulated versions of NOISEX-92's white and the pink noise are included.
Both noises are sinusoidally modulated with a frequency of $0.5~\text{Hz}$ and a modulation depth of $0.5$.
From the \url{freesound.org} database we include the last four noise types.
These are an aircraft interior noise~{\footnotesize (\url{freesound.org/s/188810})}, a babble noise~{\footnotesize (\url{freesound.org/s/88653})}, an overpassing propeller plane~{\footnotesize (\url{freesound.org/s/115387})}, traffic noise~{\footnotesize (\url{freesound.org/s/252216})} and a vacuum cleaner~{\footnotesize (\url{freesound.org/s/67421})}.

The speech material is taken from the TIMIT test set~\cite{garofolo_timit_1993} which also ensures that the speech material is different from the training.
We select 64~sentences spoken by male speakers and 64~sentences spoken by female speakers.
The 128~sentences have been recorded from 20 different speakers.
The sentences are artificially corrupted by the noises described above.
This is done in two different ways.

First, we analyze how the overall level of the input signal influences the performance of the \ac{DNN} based speech enhancement approaches.
For this, we keep the input \ac{SNR} fixed at a level of $5~\text{dB}$ and set the peak-level of all 128 sentences to different fixed values.
Each sentence hence appears with a peak level of $-40~\text{dB}$, $-24~\text{dB}$, $-18~\text{dB}$, $-12~\text{dB}$ and $-6~\text{dB}$ in the test set.
Again, the peak level is adjusted before the sentence is corrupted by the background noise.
Most levels are in the range that has also been used in the training data, whereas the peak level $-40~\text{dB}$ is an extreme case which has not been seen during training.
The results for this evaluation are discussed in Section~\ref{sec:Instrumental:Level}.

Second, we conduct an experiment where all sentences are corrupted by all noise types at input \acp{SNR} ranging from $-5~\text{dB}$ to $20~\text{dB}$.
For this experiment the peak level of each sentence randomly varied between $-26~\text{dB}$ and $-3~\text{dB}$, i.e., the same range used for training.
The results of this experiment are shown in Section~\ref{sec:Instrumental:SNR}.

\subsection{Evaluation Over the Input Level}%
\label{sec:Instrumental:Level}

\begin{figure*}[tb]
    \centering
     \begin{tikzpicture}
    \begin{customlegend}[
            legend entries={%
                conventional,
                log. periodogram,
                NAT,
                SNR-NAT,
            },
            legend columns={-1},
            legend cell align={left},
            legend style={inner sep=1pt, draw=none, font=\small},
            /tikz/every even column/.append style={column sep=0.3cm},
        ]
        \addlegendimage{generic};
        \addlegendimage{log_pery};
        \addlegendimage{log_pery_log_psdn};
        \addlegendimage{log_apost_log_aprior};
    \end{customlegend}
\end{tikzpicture}\\
    \begin{tikzpicture}
    \begin{groupplot}[%
        scale only axis,
        group style={%
            group name=instrumental,
            group size=6 by 4,
            x descriptions at=edge bottom,
            y descriptions at=edge left,
            vertical sep=1ex,
            horizontal sep=2ex,
        },
        width=0.125\textwidth,
        height=1.75cm,
        xlabel={$\text{Speech peak}~\text{[dB]}$},
        xlabel style={font=\footnotesize, yshift=0.5em},
        x tick label style={font=\footnotesize},
        y tick label style={%
            /pgf/number format/.cd,
            fixed,
            precision=2},
        grid,
        title style={anchor=south, at={(0.5, 0.85)}, font=\small},
        xmin=-40,
        xmax=-6,
        ]

        \expandafter\edef\csname limits:pesq_imp\endcsname{ymin=0.25, ymax=0.8}
        \expandafter\edef\csname limits:polqa_imp\endcsname{ymin=0.3, ymax=0.95}
        \expandafter\edef\csname limits:stoi_imp\endcsname{ymin=0, ymax=0.1}
        \expandafter\edef\csname limits:estoi_imp\endcsname{ymin=0.025, ymax=0.19}
        \expandafter\edef\csname limits:seg_ssnr\endcsname{ymin=5, ymax=15}
        \expandafter\edef\csname limits:seg_nr\endcsname{ymin=6, ymax=15}
        
        \edef\shift{xshift=1em}
        \edef\tmp{}
        \pgfplotsforeachungrouped \measure/\measurehuman in {estoi_imp/{$\Delta$ESTOI},
            polqa_imp/{$\Delta$POLQA},
            seg_ssnr/SegSSNR,
            seg_nr/SegNR} {%

            \pgfplotsforeachungrouped \dataset/\datasethuman in {
                hu_corpus_100h/{Hu corpus},
                chime3_100h/{CHiME 3},
                freesound_packs_100h/{freesound.org}} {%

                \ifx\firstrow\undefined
                \else
                \edef\datasethuman{}
                \fi

                \eappto\tmp{\noexpand\nextgroupplot[title={\datasethuman\strut},
                    ylabel={\measurehuman},
                    \csname limits:\measure\endcsname]}

                \pgfplotsforeachungrouped \feature in {log_pery_log_psdn, log_pery, log_apost_log_aprior} {%
                    \eappto\tmp{\noexpand\addplot[evalline, \feature] table[x=scaling, y=\measure] {data/instrumental/feedforward/nn_\dataset_\feature_all_fix.tsv};}
                }

                \eappto\tmp{\noexpand\addplot[evalline, generic] table[x=scaling, y=\measure] {data/instrumental/lstm/generic_all_fix.tsv};}
                \edef\measurehuman{}
            }

            \pgfplotsforeachungrouped \dataset/\datasethuman in {
                hu_corpus_100h/{Hu corpus},
                chime3_100h/{CHiME 3},
                freesound_packs_100h/{freesound.org}} {%

                \ifx\firstrow\undefined
                \else
                \edef\datasethuman{}
                \fi

                \eappto\tmp{\noexpand\nextgroupplot[title={\datasethuman\strut},
                    ylabel={\measurehuman},
                    \shift,
                    \csname limits:\measure\endcsname]}

                \pgfplotsforeachungrouped \feature in {log_pery_log_psdn, log_pery, log_apost_log_aprior} {%
                    \eappto\tmp{\noexpand\addplot[evalline, \feature] table[x=scaling, y=\measure] {data/instrumental/lstm/nn_\dataset_\feature_all_fix.tsv};}
                }

                \eappto\tmp{\noexpand\addplot[evalline, generic] table[x=scaling, y=\measure] {data/instrumental/lstm/generic_all_fix.tsv};}
                \edef\measurehuman{}
                \edef\shift{}
            }

            \def\firstrow{1}
        }
        \tmp
    \end{groupplot}

    \draw[decoration={brace, amplitude=5pt, raise=0.6cm}, decorate] (instrumental c1r1.north west) -- node[font=\small, midway, above, yshift=0.7cm] {feed-forward\strut} (instrumental c3r1.north east);
    \draw[decoration={brace, amplitude=5pt, raise=0.6cm}, decorate] (instrumental c4r1.north west) -- node[font=\small, midway, above, yshift=0.7cm] {LSTM\strut} (instrumental c6r1.north east);

    \let\firstrow=\relax
\end{tikzpicture}\\
    \caption{\acs{POLQA} improvements, \acs{ESTOI} improvements, \acs{SegSSNR} and \acs{SegNR} for feed-forward networks based (left columns) and \acs{LSTM} based (right columns) speech enhancement algorithms in dependence of the speech peak level, input feature and training dataset.
    The \acs{SNR} of the input signal is fixed at $5~\text{dB}$.
    The results of a conventional speech enhancement approach are shown for comparison.}%
        \label{fig:Results:PeakLevel}
\end{figure*}
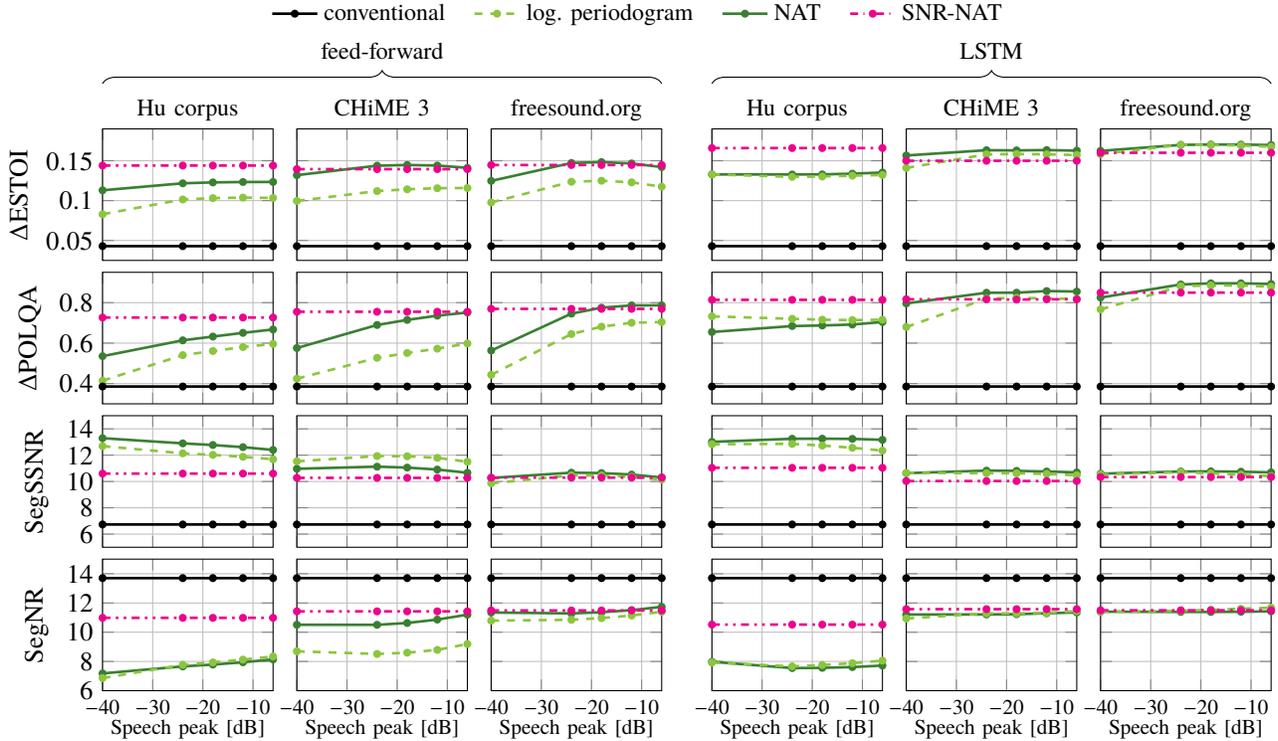

In this section, we analyze the influence of the input level on the performance of the considered speech enhancement methods.
Fig.~\ref{fig:Results:PeakLevel} depicts the results for both \ac{DNN} architectures, three different trainings sets and three different input features.
Additionally, the results for a conventional speech enhancement algorithm are shown which is based on Wiener filtering and the speech and noise \ac{PSD} estimations methods described in Section~\ref{sec:STFTEnhancement}.

The left three columns in Fig.~\ref{fig:Results:PeakLevel}, show the results for the feed-forward network.
Both \ac{ESTOI} and \ac{POLQA} show that the performance of the \ac{DNN} based enhancement schemes depends on the overall level of the input signal if periodogram features or \ac{NAT} features are used.
In general, the performance of these two enhancement schemes degrades with lower input levels.
Furthermore, the \ac{SegNR} indicates that the noise reduction becomes worse with decreasing level of the input signal.
At the same time, also the speech distortion decreases as indicated by the increasing \ac{SegSSNR}.
This indicates that the \ac{DNN} based speech enhancement algorithms reduce less noise if the periodogram and \ac{NAT} features are employed.
The scores for the conventional enhancement scheme and the proposed \ac{SNRNAT} features are virtually the same over all input levels.
This indicates that in contrast to the periodogram and \ac{NAT} features, the proposed \ac{SNRNAT} features are virtually independent of the overall level.

If \ac{LSTM} networks are employed, the level dependence becomes weaker if periodogram and \ac{NAT} features are employed.
However, this dependency is still measurable and results in lower performance if the level of the input signal drops.
Especially, for an input level of $-40~\text{dB}$ that has not been included in the training data, a small decrease in performance can be observed.
For the proposed \ac{SNRNAT} features, the performance is again virtually the same, i.e., independent of the input level.

These results also give a preview on the general performance of the considered algorithms, which we will discuss in more detail next.

\subsection{Evaluation Over the Input SNR}%
\label{sec:Instrumental:SNR}

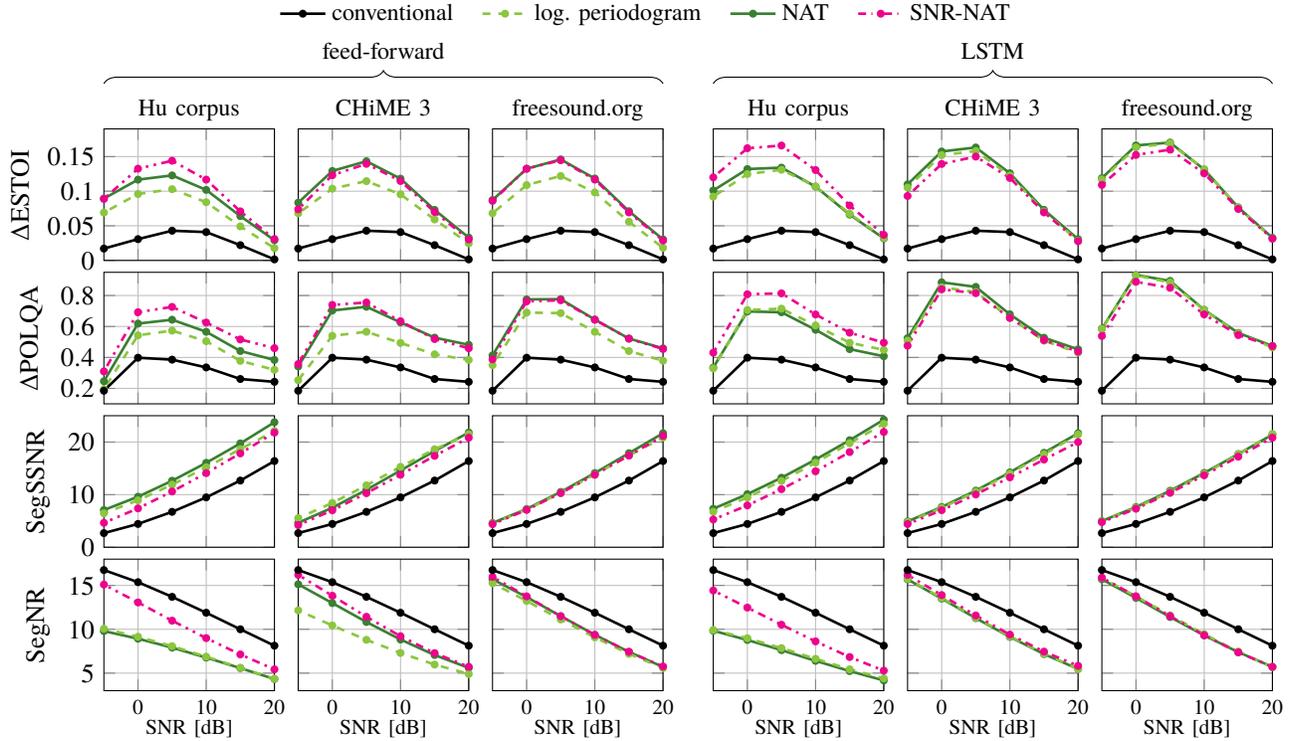
\begin{figure*}[tb]
    \centering
     \begin{tikzpicture}
    \begin{customlegend}[
            legend entries={%
                conventional,
                log. periodogram,
                NAT,
                SNR-NAT,
            },
            legend columns={-1},
            legend cell align={left},
            legend style={inner sep=1pt, draw=none, font=\small},
            /tikz/every even column/.append style={column sep=0.3cm},
        ]
        \addlegendimage{generic};
        \addlegendimage{log_pery};
        \addlegendimage{log_pery_log_psdn};
        \addlegendimage{log_apost_log_aprior};
    \end{customlegend}
\end{tikzpicture}\\
    \begin{tikzpicture}
    \begin{groupplot}[%
        scale only axis,
        group style={%
            group name=instrumental,
            group size=6 by 4,
            x descriptions at=edge bottom,
            y descriptions at=edge left,
            vertical sep=1ex,
            horizontal sep=2ex,
        },
        width=0.125\textwidth,
        height=1.75cm,
        xlabel={$\text{SNR}~\text{[dB]}$},
        xlabel style={font=\footnotesize, yshift=0.5em},
        x tick label style={font=\footnotesize},
        y tick label style={%
            /pgf/number format/.cd,
            fixed,
            precision=2},
        title style={anchor=south, at={(0.5, 0.85)}, font=\small},
        grid,
        xmin=-5,
        xmax=20,
        ]

        \expandafter\edef\csname limits:pesq_imp\endcsname{ymin=0.25, ymax=0.8}
        \expandafter\edef\csname limits:polqa_imp\endcsname{ymin=0.1, ymax=0.95}
        \expandafter\edef\csname limits:stoi_imp\endcsname{ymin=-0.025, ymax=0.125}
        \expandafter\edef\csname limits:estoi_imp\endcsname{ymin=0.0, ymax=0.19}
        \expandafter\edef\csname limits:seg_ssnr\endcsname{ymin=0, ymax=25}
        \expandafter\edef\csname limits:seg_nr\endcsname{ymin=3, ymax=18}

        \edef\shift{xshift=1em}
        \edef\tmp{}
        \pgfplotsforeachungrouped \measure/\measurehuman in {
            estoi_imp/{$\Delta$ESTOI},
            polqa_imp/{$\Delta$POLQA},
            seg_ssnr/SegSSNR,
            seg_nr/SegNR} {%

            \pgfplotsforeachungrouped \dataset/\datasethuman in {
                hu_corpus_100h/{Hu corpus},
                chime3_100h/{CHiME 3},
                freesound_packs_100h/{freesound.org}} {%

                \ifx\firstrow\undefined
                \else
                \edef\datasethuman{}
                \fi

                \eappto\tmp{\noexpand\nextgroupplot[title={\datasethuman\strut},
                    ylabel={\measurehuman},
                    \csname limits:\measure\endcsname]}

                \pgfplotsforeachungrouped \feature in {log_pery_log_psdn, log_pery, log_apost_log_aprior} {%
                    \eappto\tmp{\noexpand\addplot[evalline, \feature] table[x=snr, y=\measure] {data/instrumental/feedforward/nn_\dataset_\feature_all.tsv};}
                }

                \eappto\tmp{\noexpand\addplot[evalline, generic] table[x=snr, y=\measure] {data/instrumental/lstm/generic_all.tsv};}
                \edef\measurehuman{}
            }

            \pgfplotsforeachungrouped \dataset/\datasethuman in {
                hu_corpus_100h/{Hu corpus},
                chime3_100h/{CHiME 3},
                freesound_packs_100h/{freesound.org}} {%

                \ifx\firstrow\undefined
                \else
                \edef\datasethuman{}
                \fi

                \eappto\tmp{\noexpand\nextgroupplot[title={\datasethuman\strut},
                    ylabel={\measurehuman},
                    \shift,
                    \csname limits:\measure\endcsname]}

                \pgfplotsforeachungrouped \feature in {log_pery_log_psdn, log_pery, log_apost_log_aprior} {%
                    \eappto\tmp{\noexpand\addplot[evalline, \feature] table[x=snr, y=\measure] {data/instrumental/lstm/nn_\dataset_\feature_all.tsv};}
                }

                \eappto\tmp{\noexpand\addplot[evalline, generic] table[x=snr, y=\measure] {data/instrumental/lstm/generic_all.tsv};}
                \edef\measurehuman{}
                \edef\shift{}
            }

            \def\firstrow{1}
        }
        \tmp
    \end{groupplot}

    \draw[decoration={brace, amplitude=5pt, raise=0.6cm}, decorate] (instrumental c1r1.north west) -- node[font=\small, midway, above, yshift=0.7cm] {feed-forward\strut} (instrumental c3r1.north east);
    \draw[decoration={brace, amplitude=5pt, raise=0.6cm}, decorate] (instrumental c4r1.north west) -- node[font=\small, midway, above, yshift=0.7cm] {LSTM\strut} (instrumental c6r1.north east);

    \let\firstrow=\relax
\end{tikzpicture}\\
    \caption{\acs{POLQA} improvements, \acs{ESTOI} improvements, \acs{SegSSNR} and \ac{SegNR} for feed-forward based (left columns) and \acs{LSTM} based (right columns) speech enhancement algorithms in dependence of the input \acs{SNR}, input feature and training dataset.
        The input level is randomly varied between $-24~\text{dB}$ and $-3~\text{dB}$.
        The results of a conventional speech enhancement approach are shown for comparison.}%
    \label{fig:Results:SNR}
\end{figure*}

Similar to Section~\ref{sec:Instrumental:Level} also here, the artificially corrupted sentences are enhanced by the \ac{DNN} based speech enhancement algorithms which are trained using different features.
In contrast to Section~\ref{sec:Instrumental:Level}, we analyze the performance of the approaches in dependence of the input \ac{SNR}.
Figure~\ref{fig:Results:SNR} shows the improvements in \ac{POLQA} and \ac{ESTOI} in dependence of the input~\ac{SNR}, the employed training dataset and the employed input feature.
Further, also the absolute values for the \ac{SegSSNR} and \ac{SegNR} measures are shown again.
All measures are averaged over all noise types and again, the conventional approach is is included as a baseline.

Figure~\ref{fig:Results:SNR} shows that the performance of the \ac{DNN} based approach depends on the training dataset.
If the logarithmized periodogram or \ac{NAT} is used as input feature, the performance is considerably lower for the Hu noise corpus~\cite{hu_corpus_2005} (low amount of data) is employed.
In general, the \ac{SegNR} is lower in comparison to the \ac{SNRNAT} features and at the same time the \ac{SegSSNR} is slightly higher.
This indicates again that the \ac{DNN} based speech enhancement schemes suppress less noise, if non-robust features are used in combination with a low amount of noise training data.
Using the larger and more diverse training datasets such as CHiME~3 or the proposed collection from \url{freesound.org}, improves the performance of the periodogram and \ac{NAT} features in terms of \ac{POLQA} and \ac{ESTOI}.
If the feed-forward \ac{DNN} is considered, the performance of \ac{NAT} and \ac{SNRNAT} features are about the same, while the periodogram features yield slightly lower scores.
If the \ac{LSTM} network is used, both the periodogram and \ac{NAT} features yield slightly higher scores in \ac{POLQA} and \ac{ESTOI} than the proposed \ac{SNRNAT} features.

This analysis shows that for the proposed \ac{SNRNAT} features, the performance of the \ac{DNN} based speech enhancement approach is more robust for the considered features and training methods when only limited training data are available.
In fact, the performance using \ac{SNRNAT} is comparable for all three considered training datasets.
A dataset with limited diversity, i.e., the CHiME~3 noise corpus, has a smaller negative impact on instrumental measures.
Hence, it is possible to train well performing \acp{DNN} using a dataset with limited diversity if sufficiently powerful features are used, e.g., \ac{NAT} or \ac{SNRNAT}, or if sufficiently complex network types are used, e.g., \acp{LSTM}.
From these observations, we conclude that
\begin{enumerate*}[label=(\arabic*)]
    \item a well performing model in terms of instrumental measures can be trained independently of the feature if appropriate training data are available and
    \item that the normalization considerably improves the robustness of the \ac{DNN} based speech enhancement approach in unseen acoustic conditions if insufficient training data are available.
\end{enumerate*}

These findings are analyzed in more depth in the next section.
There, we will also show that a network that performs well in terms of averaged instrumental measures, still might struggle in unseen noise conditions.

\section{Analysis}%
\label{sec:Analysis}

In this section, we analyze and discuss the results obtained in Section~\ref{sec:InstrumentalEvaluation}.
First, we apply \ac{tSNE}~\cite{maaten_visualizing_2008} on the input features, the output of the internal of the internal representation of the second last layer and the enhanced speech signals.
The resulting plots are used to analyze the effect of the different training data and features on the generalization of \ac{DNN} based speech enhancement algorithms.
Further, we consider the masks predicted by the \ac{DNN} based speech enhancement methods for some examples taken from the test signals used in Section~\ref{sec:InstrumentalEvaluation}, i.e., for unseen noise conditions.

\subsection{T-SNE Analysis}

\begin{figure}[tb]
    \centering
    \bgroup
\tikzset{png export}
\begin{tikzpicture}[]
    \begin{groupplot}[%
        scale only axis,
        group style={
            group name=tsne,
            group size=2 by 1,
            horizontal sep=2ex,
        },
        width=0.45\linewidth,
        height=3cm,
        xlabel={},
        ylabel={},
        xticklabels={},
        yticklabels={},
        enlargelimits=0.05,
        grid,
        title style={anchor=south, yshift=-1ex, font=\footnotesize},
        ]

        \edef\tmp{}
        \pgfplotsforeachungrouped \feature/\featurehuman in {%
            log_pery_log_psdn/{NAT},
            log_apost_log_aprior/{SNR-NAT}} {%
            \eappto\tmp{\noexpand\nextgroupplot[title={\featurehuman}]}

            \pgfplotsforeachungrouped \noise in {aircraft-747-400-interior-midflight,
                                                 cocktail-amb-def,
                                                 factory1,
                                                 highway-traffic-heavy-traffic,
                                                 mod_pink,
                                                 mod_white,
                                                 vacuum_cleaner} {%
                \eappto\tmp{\noexpand\addplot[\noise, myscatter] table[x index=0, y index=1] {data/feature_analysis/scatter_feature_\feature_5.00_single_-6.00_alljournal-norandom-notalker_\noise.tsv};}
            }
        }
        \tmp
    \end{groupplot}
\end{tikzpicture}
\egroup\\
    \begin{tikzpicture}
    \begin{customlegend}[
        legend entries={%
            aircraft,
            babble,
            factory 1,
            traffic,
            vacuum cleaner,
            mod. pink,
            mod. white,
        },
        legend columns={4},
        legend cell align={left},
        legend style={inner sep=1pt, draw=none, font=\footnotesize},
        /tikz/every even column/.append style={column sep=0.1cm},
        ]
        \addlegendimage{myscatter, mark size=1.5pt, aircraft-747-400-interior-midflight};
        \addlegendimage{myscatter, mark size=1.5pt, cocktail-amb-def};
        \addlegendimage{myscatter, mark size=1.5pt, factory1};
        \addlegendimage{myscatter, mark size=1.5pt, highway-traffic-heavy-traffic};
        \addlegendimage{myscatter, mark size=1.5pt, vacuum_cleaner};
        \addlegendimage{myscatter, mark size=1.5pt, mod_pink};
        \addlegendimage{myscatter, mark size=1.5pt, mod_white};
    \end{customlegend}
\end{tikzpicture}
    \caption{T-SNE of different input features}%
    \label{fig:TSNE:Features}
\end{figure}
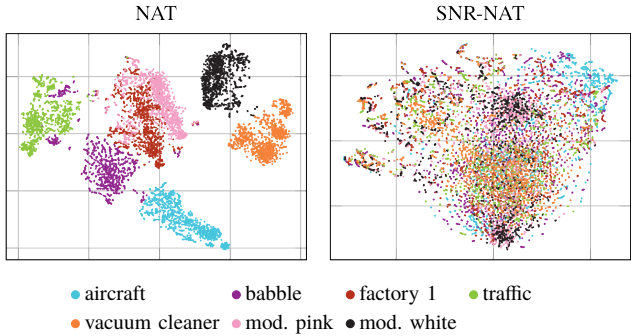

In this section, we analyze the input features \ac{NAT} and \ac{SNRNAT} and the internal representation of the feed-forward and \ac{LSTM} networks using \ac{tSNE}~\cite{maaten_visualizing_2008}.
\Ac{tSNE} is a method that embeds high dimensional vectors in a low dimensional space.
For this experiment, we extract \ac{NAT} features $\sFeatureVecNAT$ and \ac{SNRNAT} features $\sFeatureVecNorm$ from artificially corrupted speech signals, where no additional context is added to the feature vectors.
Two sentences of a male speaker and a female speaker are used where the speech signals are normalized to a maximum peak level of $-6~\text{dB}$.
After this, the sentences are corrupted by seven different background noise types at an \ac{SNR} of $5~\text{dB}$ which have not been seen during training.
\Acp{tSNE} have been obtained using the original implementation of the tree-based accelerated \ac{tSNE} algorithm described in~\cite{maaten_accelerating_2014} with a perplexity of~50 and a threshold $\theta = 0.5$.
For all datasets the algorithm was run for 1000~iterations and the \ac{tSNE} embedding converged for all datasets.
The result of the \ac{tSNE} analysis is depicted in the scatter plot shown in Fig.~\ref{fig:TSNE:Features}.
Each point in the figure corresponds to a high dimensional feature vector embedded in a low dimensional space.

As one part of the \ac{NAT} feature vector contains the noise \ac{PSD} estimate, the features highly depend on the noise type.
This can also be seen from the embeddings in Fig.~\ref{fig:TSNE:Features}, where vectors extracted from the same noise type end up in the same cluster.
The overlap between the cluster is quite small and most of the clusters are easily separable.
In contrast to the \ac{NAT} features, the embeddings for the \ac{SNRNAT} features do not show strong clusters depending on the noise type.
Instead all embeddings are mixed together and it is not easily possible to separate the data points based on the background noise type.
From this, we conclude that the \ac{SNRNAT} features are considerably less dependent on the background noise type.
This independence of the background noise may be an important property of the \ac{SNRNAT} features to train data driven speech enhancements models that generalize well to unseen noise types.

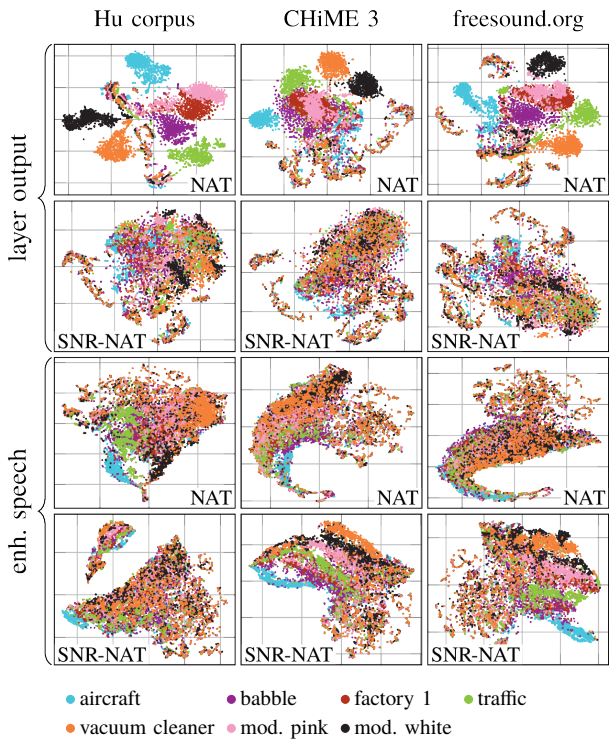
\begin{figure}[tb]
    \centering
    \bgroup%
\tikzset{png export}
\begin{tikzpicture}
    \begin{groupplot}[%
        scale only axis,
        group style={
            group name=tsne,
            group size=3 by 4,
            horizontal sep=0.5ex,
            vertical sep=0.5ex,
        },
        width=2.4cm,
        height=2cm,
        xlabel={},
        ylabel={},
        xticklabels={},
        yticklabels={},
        enlargelimits=0.05,
        grid,
        title style={anchor=south, at={(0.5, 0.9)}, font=\small},
        clip marker paths=true,
        ]

        \edef\tmp{}
        \pgfplotsforeachungrouped \set/\sethuman in {%
            hu_corpus_100h/{Hu corpus},
            chime3_100h/{CHiME 3},
            freesound_packs_100h/{freesound.org}} {%
            \eappto\tmp{\noexpand\nextgroupplot[title={\strut\sethuman}]}

            \pgfplotsforeachungrouped \noise in {aircraft-747-400-interior-midflight,
                                                 cocktail-amb-def,
                                                 factory1,
                                                 highway-traffic-heavy-traffic,
                                                 mod_pink,
                                                 mod_white,
                                                 vacuum_cleaner} {%
                \eappto\tmp{\noexpand\addplot[\noise, myscatter] table[x index=0, y index=1] {data/feature_analysis/scatter_layer_irm_\set_-2_log_pery_log_psdn_5.00_single_-6.00_alljournal-norandom-notalker_\noise.tsv};}
            }

            \eappto\tmp{\noexpand\node[anchor=south east, inner sep=1pt, font=\noexpand\footnotesize, fill=white] at (rel axis cs:1.00, 0.00) {NAT};}
        }

        \pgfplotsforeachungrouped \set in {%
            hu_corpus_100h,
            chime3_100h,
            freesound_packs_100h} {%
            \eappto\tmp{\noexpand\nextgroupplot}

            \pgfplotsforeachungrouped \noise in {aircraft-747-400-interior-midflight,
                                                 cocktail-amb-def,
                                                 factory1,
                                                 highway-traffic-heavy-traffic,
                                                 mod_pink,
                                                 mod_white,
                                                 vacuum_cleaner} {%
                \eappto\tmp{\noexpand\addplot[\noise, myscatter] table[x index=0, y index=1] {data/feature_analysis/scatter_layer_irm_\set_-2_log_apost_log_aprior_5.00_single_-6.00_alljournal-norandom-notalker_\noise.tsv};}
            }

            \eappto\tmp{\noexpand\node[anchor=south west, inner sep=1pt, font=\noexpand\footnotesize, fill=white] at (rel axis cs:0.00, 0.00) {SNR-NAT};}
        }

        \pgfplotsforeachungrouped \set in {%
            hu_corpus_100h,
            chime3_100h,
            freesound_packs_100h} {%
            \eappto\tmp{\noexpand\nextgroupplot}

            \pgfplotsforeachungrouped \noise in {aircraft-747-400-interior-midflight,
                                                 cocktail-amb-def,
                                                 factory1,
                                                 highway-traffic-heavy-traffic,
                                                 mod_pink,
                                                 mod_white,
                                                 vacuum_cleaner} {%
                \eappto\tmp{\noexpand\addplot[\noise, myscatter] table[x index=0, y index=1] {data/feature_analysis/scatter_enhancement_irm_\set_log_pery_log_psdn_5.00_single_-6.00_alljournal-norandom-notalker_\noise.tsv};}
            }

            \eappto\tmp{\noexpand\node[anchor=south east, inner sep=1pt, font=\noexpand\footnotesize, fill=white] at (rel axis cs:1.00, 0.00) {NAT};}
        }

        \pgfplotsforeachungrouped \set in {%
            hu_corpus_100h,
            chime3_100h,
            freesound_packs_100h} {%
            \eappto\tmp{\noexpand\nextgroupplot}

            \pgfplotsforeachungrouped \noise in {aircraft-747-400-interior-midflight,
                                                 cocktail-amb-def,
                                                 factory1,
                                                 highway-traffic-heavy-traffic,
                                                 mod_pink,
                                                 mod_white,
                                                 vacuum_cleaner} {%
                \eappto\tmp{\noexpand\addplot[\noise, myscatter] table[x index=0, y index=1] {data/feature_analysis/scatter_enhancement_irm_\set_log_apost_log_aprior_5.00_single_-6.00_alljournal-norandom-notalker_\noise.tsv};}
            }

            \eappto\tmp{\noexpand\node[anchor=south west, inner sep=1pt, font=\noexpand\footnotesize, fill=white] at (rel axis cs:0.00, 0.00) {SNR-NAT};}
        }

        \tmp
    \end{groupplot}

    \draw[-, decoration={brace, mirror, amplitude=5pt, raise=1pt}, decorate] (tsne c1r1.north west) -- node[midway, left, xshift=-5pt] {\rotatebox{90}{layer output}} (tsne c1r2.south west);

    \draw[-, decoration={brace, mirror, amplitude=5pt, raise=1pt}, decorate] (tsne c1r3.north west) -- node[midway, left, xshift=-5pt] {\rotatebox{90}{enh.\ speech}} (tsne c1r4.south west);
\end{tikzpicture}
\egroup%
\\
    \begin{tikzpicture}
    \begin{customlegend}[
        legend entries={%
            aircraft,
            babble,
            factory 1,
            traffic,
            vacuum cleaner,
            mod. pink,
            mod. white,
        },
        legend columns={4},
        legend cell align={left},
        legend style={inner sep=1pt, draw=none, font=\footnotesize},
        /tikz/every even column/.append style={column sep=0.1cm},
        ]
        \addlegendimage{myscatter, mark size=1.5pt, aircraft-747-400-interior-midflight};
        \addlegendimage{myscatter, mark size=1.5pt, cocktail-amb-def};
        \addlegendimage{myscatter, mark size=1.5pt, factory1};
        \addlegendimage{myscatter, mark size=1.5pt, highway-traffic-heavy-traffic};
        \addlegendimage{myscatter, mark size=1.5pt, vacuum_cleaner};
        \addlegendimage{myscatter, mark size=1.5pt, mod_pink};
        \addlegendimage{myscatter, mark size=1.5pt, mod_white};
    \end{customlegend}
\end{tikzpicture}
    \caption{T-SNE for feed-forward networks}%
    \label{fig:TSNE:Feedforward}
\end{figure}

\begin{figure}[tb]
    \centering
    \bgroup%
\tikzset{png export}
\begin{tikzpicture}
    \begin{groupplot}[%
        scale only axis,
        group style={
            group name=tsne,
            group size=3 by 4,
            horizontal sep=0.5ex,
            vertical sep=0.5ex,
        },
        width=2.4cm,
        height=2cm,
        xlabel={},
        ylabel={},
        xticklabels={},
        yticklabels={},
        enlargelimits=0.05,
        grid,
        title style={anchor=south, at={(0.5, 0.9)}, font=\small},
        clip marker paths=true,
        ]

        \edef\tmp{}
        \pgfplotsforeachungrouped \set/\sethuman in {%
            hu_corpus_100h/{Hu corpus},
            chime3_100h/{CHiME 3},
            freesound_packs_100h/{freesound.org}} {%
            \eappto\tmp{\noexpand\nextgroupplot[title={\strut\sethuman}]}

            \pgfplotsforeachungrouped \noise in {aircraft-747-400-interior-midflight,
                                                 cocktail-amb-def,
                                                 factory1,
                                                 highway-traffic-heavy-traffic,
                                                 mod_pink,
                                                 mod_white,
                                                 vacuum_cleaner} {%
                \eappto\tmp{\noexpand\addplot[\noise, myscatter] table[x index=0, y index=1] {data/feature_analysis/scatter_layer_lstm_\set_-2_log_pery_log_psdn_5.00_single_-6.00_alljournal-norandom-notalker_\noise.tsv};}
            }

            \eappto\tmp{\noexpand\node[anchor=south east, inner sep=1pt, font=\noexpand\footnotesize, fill=white] at (rel axis cs:1.00, 0.00) {NAT};}
        }

        \pgfplotsforeachungrouped \set in {%
            hu_corpus_100h,
            chime3_100h,
            freesound_packs_100h} {%
            \eappto\tmp{\noexpand\nextgroupplot}

            \pgfplotsforeachungrouped \noise in {aircraft-747-400-interior-midflight,
                                                 cocktail-amb-def,
                                                 factory1,
                                                 highway-traffic-heavy-traffic,
                                                 mod_pink,
                                                 mod_white,
                                                 vacuum_cleaner} {%
                \eappto\tmp{\noexpand\addplot[\noise, myscatter] table[x index=0, y index=1] {data/feature_analysis/scatter_layer_lstm_\set_-2_log_apost_log_aprior_5.00_single_-6.00_alljournal-norandom-notalker_\noise.tsv};}
            }

            \eappto\tmp{\noexpand\node[anchor=south west, inner sep=1pt, font=\noexpand\footnotesize, fill=white] at (rel axis cs:0.00, 0.00) {SNR-NAT};}
        }

        \pgfplotsforeachungrouped \set in {%
            hu_corpus_100h,
            chime3_100h,
            freesound_packs_100h} {%
            \eappto\tmp{\noexpand\nextgroupplot}

            \pgfplotsforeachungrouped \noise in {aircraft-747-400-interior-midflight,
                                                 cocktail-amb-def,
                                                 factory1,
                                                 highway-traffic-heavy-traffic,
                                                 mod_pink,
                                                 mod_white,
                                                 vacuum_cleaner} {%
                \eappto\tmp{\noexpand\addplot[\noise, myscatter] table[x index=0, y index=1] {data/feature_analysis/scatter_enhancement_lstm_\set_log_pery_log_psdn_5.00_single_-6.00_alljournal-norandom-notalker_\noise.tsv};}
            }

            \eappto\tmp{\noexpand\node[anchor=south west, inner sep=1pt, font=\noexpand\footnotesize, fill=white] at (rel axis cs:0.00, 0.00) {NAT};}
        }

        \pgfplotsforeachungrouped \set in {%
            hu_corpus_100h,
            chime3_100h,
            freesound_packs_100h} {%
            \eappto\tmp{\noexpand\nextgroupplot}

            \pgfplotsforeachungrouped \noise in {aircraft-747-400-interior-midflight,
                                                 cocktail-amb-def,
                                                 factory1,
                                                 highway-traffic-heavy-traffic,
                                                 mod_pink,
                                                 mod_white,
                                                 vacuum_cleaner} {%
                \eappto\tmp{\noexpand\addplot[\noise, myscatter] table[x index=0, y index=1] {data/feature_analysis/scatter_enhancement_lstm_\set_log_apost_log_aprior_5.00_single_-6.00_alljournal-norandom-notalker_\noise.tsv};}
            }

            \eappto\tmp{\noexpand\node[anchor=south east, inner sep=1pt, font=\noexpand\footnotesize, fill=white] at (rel axis cs:1.00, 0.00) {SNR-NAT};}
        }

        \tmp
    \end{groupplot}

    \draw[-, decoration={brace, mirror, amplitude=5pt, raise=1pt}, decorate] (tsne c1r1.north west) -- node[midway, left, xshift=-5pt] {\rotatebox{90}{layer output}} (tsne c1r2.south west);

    \draw[-, decoration={brace, mirror, amplitude=5pt, raise=1pt}, decorate] (tsne c1r3.north west) -- node[midway, left, xshift=-5pt] {\rotatebox{90}{enh.\ speech}} (tsne c1r4.south west);
\end{tikzpicture}
\egroup%
\\
    \begin{tikzpicture}
    \begin{customlegend}[
        legend entries={%
            aircraft,
            babble,
            factory 1,
            traffic,
            vacuum cleaner,
            mod. pink,
            mod. white,
        },
        legend columns={4},
        legend cell align={left},
        legend style={inner sep=1pt, draw=none, font=\footnotesize},
        /tikz/every even column/.append style={column sep=0.1cm},
        ]
        \addlegendimage{myscatter, mark size=1.5pt, aircraft-747-400-interior-midflight};
        \addlegendimage{myscatter, mark size=1.5pt, cocktail-amb-def};
        \addlegendimage{myscatter, mark size=1.5pt, factory1};
        \addlegendimage{myscatter, mark size=1.5pt, highway-traffic-heavy-traffic};
        \addlegendimage{myscatter, mark size=1.5pt, vacuum_cleaner};
        \addlegendimage{myscatter, mark size=1.5pt, mod_pink};
        \addlegendimage{myscatter, mark size=1.5pt, mod_white};
    \end{customlegend}
\end{tikzpicture}
    \caption{T-SNE for \acs{LSTM} networks}%
    \label{fig:TSNE:LSTM}
\end{figure}
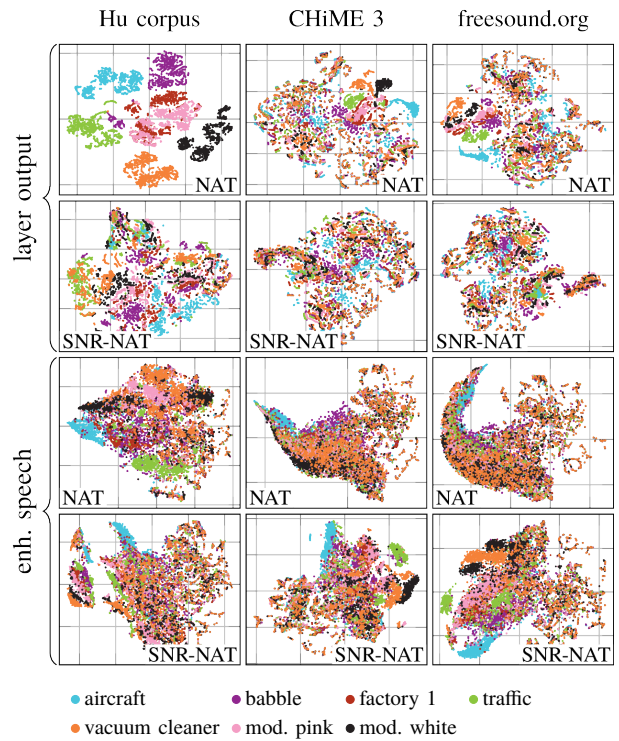

The same audio files have been used to extract the embeddings shown in Fig.~\ref{fig:TSNE:Feedforward} and Fig.~\ref{fig:TSNE:LSTM}.
In contrast to Fig.~\ref{fig:TSNE:Features}, the embeddings have been computed from the second last layer of a trained network, i.e., an internal representation of the neural network, and the absolute magnitudes of the enhanced speech coefficients.
Fig.~\ref{fig:TSNE:Feedforward} shows the results for the feed-forward network, while Fig.~\ref{fig:TSNE:LSTM} shows the results for the \ac{LSTM} network.
In both figures, the upper two rows show the embeddings for the internal representation and the last two rows depict the embedding for the enhanced speech signal.
Furthermore, both figures depict the embeddings for different training data sets, which are shown in the different columns.

The first row of Fig.~\ref{fig:TSNE:Feedforward} shows the embeddings of the internal representations of the feed-forward network, which has been trained using \ac{NAT} features.
Interestingly, also the deep internal representation of the network appears to be dependent on the noise type as indicated by clusters that form for each noise type similar to Fig.~\ref{fig:TSNE:Features}.
Contrarily, the internal representation that is obtained when using the \ac{SNRNAT} features, depicted in the second row of Fig.~\ref{fig:TSNE:Feedforward}, does not show this behaviors and no clustering can be observed.
These observations can be made for all training data sets.

The third and the fourth row in Fig.~\ref{fig:TSNE:Feedforward} shows the embeddings for the estimated clean speech coefficients.
A strong clustering of the enhanced speech coefficients with respect to the noise type is thought to be an undesired effect because the estimated clean speech coefficient should be independent of the background noise.
Still, some dependence on the noise type is to be expected as the noise is only attenuated and not completely suppressed in our experiments.
The embeddings of estimated speech coefficients shown in the third row, which result when using the \ac{NAT} features, show some dependence on the training data.
Some noise type dependent clustering is observable if the Hu corpus~\cite{hu_corpus_2005} is employed for training, but the data points are much more mixed as compared to the layer output.
Still, the somewhat stronger dependence of the estimated clean speech coefficients on the noise type might explain the lower scores in \ac{ESTOI} and \ac{POLQA} shown in Fig.~\ref{fig:Results:SNR}.
The embeddings for the CHiME~3 dataset~\cite{barker_third_2017} and the \url{freesound.org} dataset are quite similar.
In general, no obvious dependence on the noise type is visible and the scores obtained for both training datasets are similarly high.
The embedding of the enhanced speech coefficients obtained when using the \ac{SNRNAT} features are generally mixed.
Interestingly, the embeddings seem to be more clustered when using the CHiME~3 dataset~\cite{barker_third_2017} and the \url{freesound.org} dataset for training.
However, this type of clustering does not appear to have a relationship to the \ac{ESTOI} and \ac{POLQA} scores shown in Fig.~\ref{fig:Results:SNR}.

Fig.~\ref{fig:TSNE:LSTM} shows similar embeddings as Fig.~\ref{fig:TSNE:Feedforward} but here the \ac{LSTM} network has been used instead of the feed-forward network.
One of the most clearly visible differences to Fig.~\ref{fig:TSNE:Feedforward} is shown in the first row.
The embeddings of the internal representations are less clustered if the network is trained on the CHiME~3 dataset~\cite{barker_third_2017} or the \url{freesound.org} dataset using \ac{NAT} features.
If the Hu corpus is used in combination with \ac{NAT} features for training, the internal representation is still clustered similar to the feed-forward network.
This indicates that the \ac{LSTM} networks are able to find an internal representation that is more independent of the noise type.
Looking at the embeddings of the estimated clean speech coefficients that have been obtained using \ac{NAT} features, it appears that these are more clustered if the network is trained on the Hu corpus~\cite{hu_corpus_2005} in comparison to the other training datasets.
Again, this is in line with the observation that the network yields the lowest scores in \ac{ESTOI} and \ac{POLQA} as shown in Fig.~\ref{fig:Results:SNR}.
For the \ac{SNRNAT} features, the observations obtained from the \ac{LSTM} network are similar to the feed-forward network.

From these observations, we follow that \ac{NAT} features are noise dependent.
Further, their usage also leads to a noise dependent internal representation for feed-forward networks.
If an \ac{LSTM} network is used and the training data is sufficiently diverse, i.e., the CHiME~3 corpus~\cite{barker_third_2017} or the \url{freesound.org} is used, a noise independent internal representation can be learnt.
Contrarily, the proposed \ac{SNRNAT} features are independent of the noise type and hence, also lead to a internal representation that is independent of the background noise.
This can be observed for both the employed feed-forward network and \ac{LSTM} network.
Therefore, the \ac{SNRNAT} are more robust to issues in the design of the training data and applications with small amounts of data available.
Further, \ac{SNRNAT} features are scale-invariant.

\subsection{Predicted Masks in Unseen Noise Conditions}

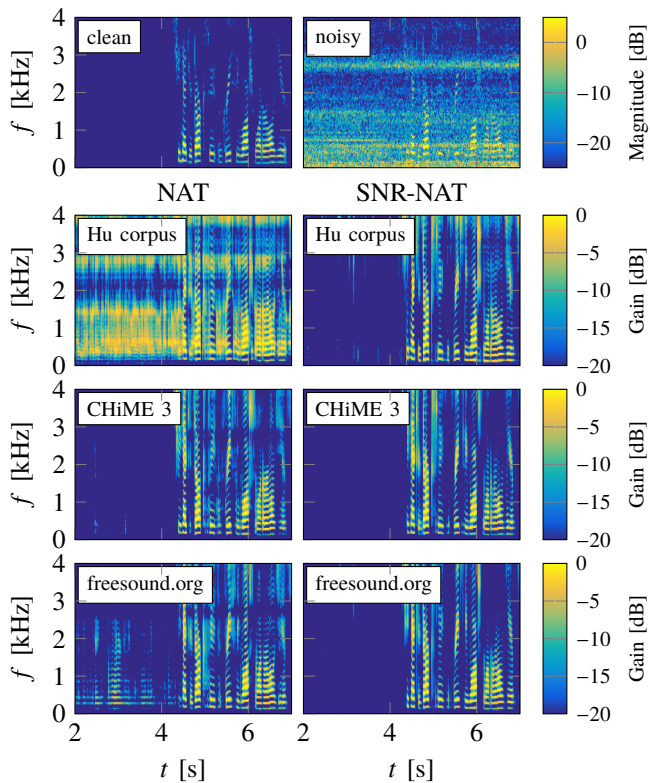
\begin{figure}[tb]
    \centering
    \begin{tikzpicture}
    \begin{groupplot}[%
        axis on top,
        scale only axis,
        group style={
            group size=2 by 4,
            x descriptions at=edge bottom,
            y descriptions at=edge left,
            horizontal sep=1ex,
            vertical sep=2ex,
        },
        width=0.325\linewidth,
        height=2cm,
        xlabel={$t~\text{[s]}$},
        ylabel={$f~\text{[kHz]}$},
        title style={yshift=-1.75ex},
        xmin=2,
        xmax=7,
        ymin=0,
        ymax=4,
        colormap name={parula},
        colorbar style={width=2ex, font=\footnotesize, ylabel={Gain [dB]}},
        ]

        \edef\noisetype{f16}
        \edef\tmp{}

        \let\firstrow\undefined
        \expandafter\edef\csname style:NAT\endcsname{}
        \expandafter\edef\csname style:SNR-NAT\endcsname{colorbar}

        \eappto\tmp{
            \noexpand\nextgroupplot[point meta min=-25, point meta max=5]
            \noexpand\getImage{data/examples/clean_vacuum_cleaner_0.tsv}
            \noexpand\node[fill=white, rectangle, draw, anchor=north west, font=\noexpand\footnotesize] at($(rel axis cs: 0, 1.0) + (1pt, -1pt)$) {clean};

            \noexpand\nextgroupplot[point meta min=-25, point meta max=5, colorbar, colorbar style={ylabel={Magnitude [dB]}}]
            \noexpand\getImage{data/examples/noisy_f16_0.tsv}
            \noexpand\node[fill=white, rectangle, draw, anchor=north west, font=\noexpand\footnotesize] at($(rel axis cs: 0, 1.0) + (1pt, -1pt)$) {noisy};
        }

        \pgfplotsforeachungrouped \dataset/\datasethuman in {hu_corpus_100h/Hu corpus, chime3_100h/CHiME 3, freesound_packs_100h/freesound.org} {%
            \pgfplotsforeachungrouped \feature/\featurehuman in {log_pery:0:0_log_psdn:0:0/NAT, log_apost:0:0_log_aprior:0:0/{SNR-NAT}} {%
                \ifx\firstrow\undefined
                \edef\title{\featurehuman}
                \edef\yshift{yshift=-2ex}
                \else
                \edef\title{}
                \edef\yshift{}
                \fi

                \edef\additionalstyle{\csname style:\featurehuman\endcsname}

                \eappto\tmp{\noexpand\nextgroupplot[title={\noexpand\strut\title}, \additionalstyle, \yshift, point meta min=-20, point meta max=0]}
                \eappto\tmp{\noexpand\getImage{data/examples/\feature_\dataset_\noisetype_0.tsv}}
                \eappto\tmp{\noexpand\node[fill=white, rectangle, draw, anchor=north west, font=\noexpand\footnotesize] at($(rel axis cs: 0, 1) + (1pt, -1pt)$) {\datasethuman};}
            }
            \edef\firstrow{}
        }

        \tmp

    \end{groupplot}
\end{tikzpicture}
    \caption{Masks $\sGainIIRMEst$ obtained for a speech signal corrupted by F16 noise at $0~\text{dB}$ \acs{SNR} using the \acs{LSTM} based speech enhancement network for three different training data sets and two different input features.
    The first row shows the clean and the noisy data.}%
    \label{fig:Examples:F16}
\end{figure}

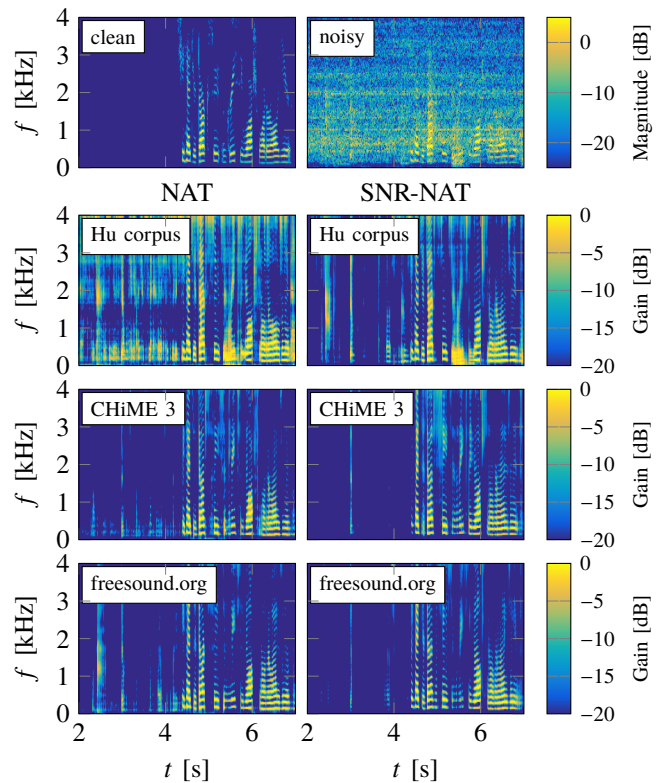
\begin{figure}[tb]
    \centering
    \begin{tikzpicture}
    \begin{groupplot}[%
        axis on top,
        scale only axis,
        group style={
            group size=2 by 4,
            x descriptions at=edge bottom,
            y descriptions at=edge left,
            horizontal sep=1ex,
            vertical sep=2ex,
        },
        width=0.325\linewidth,
        height=2cm,
        xlabel={$t~\text{[s]}$},
        ylabel={$f~\text{[kHz]}$},
        title style={yshift=-1.75ex},
        xmin=2,
        xmax=7,
        ymin=0,
        ymax=4,
        colormap name={parula},
        colorbar style={width=2ex, font=\footnotesize, ylabel={Gain [dB]}},
        ]

        \edef\noisetype{vacuum_cleaner}
        \edef\tmp{}

        \let\firstrow\undefined
        \expandafter\edef\csname style:NAT\endcsname{}
        \expandafter\edef\csname style:SNR-NAT\endcsname{colorbar}

        \eappto\tmp{
            \noexpand\nextgroupplot[point meta min=-25, point meta max=5]
            \noexpand\getImage{data/examples/clean_vacuum_cleaner_0.tsv}
            \noexpand\node[fill=white, rectangle, draw, anchor=north west, font=\noexpand\footnotesize] at($(rel axis cs: 0, 1.0) + (1pt, -1pt)$) {clean};

            \noexpand\nextgroupplot[point meta min=-25, point meta max=5, colorbar, colorbar style={ylabel={Magnitude [dB]}}]
            \noexpand\getImage{data/examples/noisy_vacuum_cleaner_0.tsv}
            \noexpand\node[fill=white, rectangle, draw, anchor=north west, font=\noexpand\footnotesize] at($(rel axis cs: 0, 1.0) + (1pt, -1pt)$) {noisy};
        }

        \pgfplotsforeachungrouped \dataset/\datasethuman in {hu_corpus_100h/Hu corpus, chime3_100h/CHiME 3, freesound_packs_100h/freesound.org} {%
            \pgfplotsforeachungrouped \feature/\featurehuman in {log_pery:0:0_log_psdn:0:0/NAT, log_apost:0:0_log_aprior:0:0/{SNR-NAT}} {%
                \ifx\firstrow\undefined
                \edef\title{\featurehuman}
                \edef\yshift{yshift=-2ex}
                \else
                \edef\title{}
                \edef\yshift{}
                \fi

                \edef\additionalstyle{\csname style:\featurehuman\endcsname}

                \eappto\tmp{\noexpand\nextgroupplot[title={\noexpand\strut\title}, \additionalstyle, \yshift, point meta min=-20, point meta max=0]}
                \eappto\tmp{\noexpand\getImage{data/examples/\feature_\dataset_\noisetype_0.tsv}}
                \eappto\tmp{\noexpand\node[fill=white, rectangle, draw, anchor=north west, font=\noexpand\footnotesize] at($(rel axis cs: 0, 1.0) + (1pt, -1pt)$) {\datasethuman};}
            }
            \edef\firstrow{}
        }

        \tmp

    \end{groupplot}
\end{tikzpicture}
    \caption{Masks $\sGainIIRMEst$ obtained for a speech signal corrupted by vacuum cleaner noise at $0~\text{dB}$ \acs{SNR} using the \acs{LSTM} based speech enhancement network for three different training data sets and two different input features.
    The first row shows the clean and the noisy data.}%
    \label{fig:Examples:Vacuum}
\end{figure}

In this section, examples are used to demonstrate the advantages and disadvantages of the different input features.
For this, a single speech signal is corrupted by F16 noise and vacuum cleaner noise at an \ac{SNR} of $0~\text{dB}$.
The first four seconds of the noisy input signals contain only noise.
The signal is processed using the \ac{LSTM} based speech enhancement networks that have been trained using the previously described input features.
Figure~\ref{fig:Examples:F16} and~\ref{fig:Examples:Vacuum} shows the resulting masks that result from the respective networks.

Both noise types are relatively stationary and should not pose a problem to the speech enhancement algorithms.
But the masks clearly show that these noise types can be challenging for the \ac{DNN} based speech enhancement algorithms.
If the Hu corpus and \ac{NAT} features are used for training, the resulting network is unable to cope with these unseen noise types.
Instead of reducing the noise, the mask shows values close to $0~\text{dB}$ at random positions in the noise only region at the beginning of the signals.
This behavior is in line with the reduced noise reduction and increased \acs{SegSSNR}, which has been observed in Section~\ref{sec:InstrumentalEvaluation} for the networks trained with the Hu corpus.
The random behavior of the mask in noise regions is clearly audible and degrades the signal quality in noise only regions drastically.
Using the proposed \ac{SNRNAT} features, the network correctly reduces the noise in regions where only noise is present.
Consequently, a lot less artifacts and random openings of the mask are observable in the enhanced signal and, as a result, the quality of the processed signal is considerably better with the proposed \ac{SNRNAT} features.

Even though the behavior is not as extreme as for the Hu corpus, such errors can also be observed for more complex training datasets when the \ac{NAT} features are used.
Despite the high scores in \ac{POLQA} and \ac{ESTOI} for networks trained on the CHiME~3 dataset and the proposed \url{freesound.org} dataset with \ac{NAT} features, there are still artifacts observable in the noise only regions.
If trained on the \url{freesound.org} corpus, the network modulates speech-like sounds into the noise as shown in Figure~\ref{fig:Examples:F16} if \ac{NAT} features are used.
In the vacuum cleaner noise shown in Figure~\ref{fig:Examples:Vacuum}, several spots can be observed in lower frequencies for the mask of the network trained on the CHiME~3 dataset despite only noise being present.
Even though the modulations are less extreme, it can be verified using sound examples, that the observed artifacts are still audible.
Contrarily, the background noise is much smoother when the proposed \ac{SNRNAT} features are employed leading to robust results in practical use cases.

The examples used in Fig.~\ref{fig:Examples:F16} and Fig.~\ref{fig:Examples:Vacuum} are also available as sound examples\footnote{\sURL}.
Additionally, the website contains excerpts of the real evaluation set of the CHiME~3 challenge~\cite{barker_third_2015, barker_third_2017} and a video recording from our lab, which have been processed using the algorithms considered here.
In contrast to the signals used here, the CHiME~3 and the lab video examples on the website are real recordings, i.e., speech and noise have not been artificially mixed.
The sound examples show that the conclusions from Section~\ref{sec:InstrumentalEvaluation} and this section also apply to real signals.

\section{Conclusions}%
\label{sec:Conclusions}

In this paper, we addressed the generalization of \ac{DNN} based speech enhancement algorithms to unseen noise.
For this, we analyzed the impact of different input features, different training data sets and different network architectures.
Furthermore, we propose \ac{SNRNAT} features and a noise dataset based on sounds collected from \url{freesound.org} to improve the robustness of \ac{DNN} based speech enhancement algorithms.

Our findings indicate that diverse training data that are severely limited in size, e.g., the Hu corpus~\cite{hu_corpus_2005}, may severly limit the generalization of speech enhancement networks, especially if simplistic features are used.
Contrarily, a large dataset results in a better generalization, even if the diversity is limited, as for the CHiME~3 corpus~\cite{barker_third_2017} which consists of only four noise environments. However, we showed that for noise sounds that do not follow the typical spectral envelope of the CHiME~3 noise environments, like a vacuum cleaner or F16 noise, the performance may still be unsatisfactory.
Generally, a large and diverse dataset such as the proposed dataset constructed from sounds from \url{freesound.org} is required to train a speech enhancement network that generalizes well to unseen conditions.

Further, we show that the proposed \ac{SNRNAT} features lead to more robust networks even if training data limited in size and diversity are used.
This result is confirmed also by an analysis via \ac{tSNE} which shows less noise-specific clustering for \ac{SNRNAT} features than for periodogram or \ac{NAT} features.
The \ac{tSNE} analysis shows that the \ac{LSTM} network is able to disentangle the dependency on the noise type of periodgram and \ac{NAT} features if sufficient training data are available.
Using \ac{SNRNAT} features as input feature, however, the internal representation becomes noise independent also for less complex network types such as the feed-forward network.
The examples in Fig.~\ref{fig:Examples:F16} and Fig.~\ref{fig:Examples:Vacuum} further show, that also \ac{LSTM} networks benefit from \ac{SNRNAT} features especially in noise only regions and ensure that also unseen noises are correctly suppressed.

\ifCLASSOPTIONcaptionsoff
  \newpage
\fi

\bibliographystyle{IEEEtran}
\bibliography{bibliography/Dissertation.bib}

\end{document}